\documentclass[preprint,12pt,twocolumb]{elsarticle}
\usepackage[latin1]{inputenc}
\usepackage{amsmath}
\usepackage{booktabs,array,multirow}
\usepackage{amsfonts}
\usepackage{blindtext}
\usepackage{amssymb}
\usepackage{graphicx}
\usepackage{chngpage}
\usepackage{amsmath}
\usepackage{subcaption}
\usepackage{siunitx}
\usepackage{float}
\DeclareSIUnit\gauss{G}

\newcommand{\RomanNumeralCaps}[1]{\MakeUppercase{\romannumeral #1}}

\usepackage{makecell}
\usepackage{xcolor}
\usepackage{latexsym}
\usepackage{textcomp}

\usepackage{tabulary}

\usepackage{chemformula} 
\usepackage[T1]{fontenc} 
\usepackage[labelfont=bf]{caption}
\newcolumntype{M}[1]{>{\centering\arraybackslash}m{#1}}

\usepackage{xr}

\journal{Carbon}

\begin{document}
	
\begin{frontmatter}

\title{Efficient Conversion of Nitrogen to Nitrogen-Vacancy Centers in Diamond Particles with High-Temperature Electron Irradiation}

\author[a]{Yuliya Mindarava\corref{mycorrespondingauthor}}
\cortext[mycorrespondingauthor]{Corresponding author. Tel: +4915203294898}
\ead{yuliya.mindarava@uni-ulm.de}
\author[a]{ R\'{e}mi Blinder\corref{mycorrespondingauthor1}}
\cortext[mycorrespondingauthor1]{Corresponding author. Tel: +497315023728}
\ead{remi.blinder@uni-ulm.de}
\author[b]{ Christian Laube}
\author[b]{ Wolfgang Knolle}
\author[b]{Bernd Abel}
\author[c]{Christian Jentgens}
\author[d]{Junichi Isoya}
\author[e]{Jochen Scheuer}
\author[a]{Johannes Lang}
\author[e]{Ilai Schwartz}
\author[f]{Boris Naydenov}
\author[a,g]{Fedor Jelezko}
\address[a]{Institute for Quantum Optics, Ulm University, Ulm 89081, Germany}

\address[b]{Department of Functional Surfaces, Leibniz Institute of Surface Engineering, Leipzig 04318, Germany}
\address[c]{Microdiamant AG, Kreuzlingerstrasse 1, CH-8574 Lengwil, Switzerland}
\address[d]{Faculty of Pure and Applied Sciences, University of Tsukuba, Tsukuba, 305-8573, Japan}
\address[e]{NVision Imaging Technologies GmbH, Ulm 89081, Germany}
\address[f]{Institute for Nanospectroscopy, Helmholtz-Zentrum Berlin f\"{u}r Materialen und Energie GmbH, Berlin 14109, Germany}
\address[g]{Centre for Integrated Quantum Science and Technology (IQST), Ulm 89081, Germany}


\begin{abstract}
	Fluorescent nanodiamonds containing negatively-charged nitrogen-vacancy (NV$^-$) centers are promising for a wide range of applications, such as for sensing, as fluorescence biomarkers, or to hyperpolarize nuclear spins. NV$^-$ centers are formed from substitutional nitrogen (P1 centers) defects and vacancies in the diamond lattice.  
	Maximizing the concentration of NVs is most beneficial, which justifies the search for methods with a high yield of conversion from P1 to NV$^-$. 
	We report here the characterization of  surface cleaned fluorescent micro- and nanodiamonds, obtained by irradiation of commercial diamond powder with high-energy (10 MeV) electrons and simultaneous annealing at 800 $^{\circ}$C. 
	Using this technique and increasing the irradiation dose, we demonstrate the creation of NV$^-$ with up to 25~\% conversion yield.
   Finally, we monitor the creation of irradiation-induced spin-1 defects in microdiamond particles, which we associate with W16 and W33 centers, and investigate the effects of irradiation dose and particle size on the coherence time of NV$^-$.

\end{abstract}

\end{frontmatter}


\section{Introduction}
The negatively-charged Nitrogen-Vacancy (NV$^-$) center, a point defect in diamond, has been established as a promising color center with unique properties, such as the ability to be polarized in the $m_s = 0$ spin-state by illumination with green light, a long spin coherence time at room temperature, and bright fluorescence without photobleaching \cite{Doherty13}. Owing to their optical and chemical properties, as well as exceptional biocompatibility \cite{Biocompatibility3}, fluorescent nanodiamonds (FNDs) with high NV$^-$ concentration are promising candidates for the synthesis of fluorescence markers \cite{Biocompatibility,NDbioapplications}, sensing probes \cite{Wu_Fedor,NDforBiology} and polarized MRI tracers \cite{MRI,Ajoy18}. The brightness of FNDs depends on the average number of light-emitting color centers per particle and therefore requires a significant concentration of NV$^-$.

The creation of NV$^-$ requires the presence of two types of impurities in the lattice, substitutional nitrogen defects (P1 centers), and vacancies. A single vacancy can recombine with a P1 center.  Provided the presence of an electron donor, which can be another P1 center, such a recombination can lead to the formation of  NV$^-$. In conventional methods, the vacancies are created by irradiation at room temperature with high energy electrons \cite{Laube2019,Electron_irrad,Laube2017}, protons \cite{Proton_irrad} or gamma rays \cite{gamma_irr}. At room temperature, however, the created vacancies remain at a fixed position in the diamond lattice.  In order to induce the mobility of vacancies towards P1 centers, an additional step of annealing is required, which practically  consists in heating up the sample above 800$^\circ$C. We label this method the room temperature (RT) irradiation technique.
  
A high NV$^-$ concentration can be achieved by irradiating nanodiamonds (NDs) with a high nitrogen content.
However, single nitrogen defects in diamonds are also a major source for spin decoherence \cite{barry2019}, photoluminescence quenching \cite{PLint_vs_P1conc,PL_vs_P1conc, su2013}, and $^{13}$C nuclear spin relaxation \cite{reynhardt2001, ajoy2019T1}. Therefore, a high residual P1 concentration reduces the nanodiamonds suitability for applications where the named properties  are crucial. 
As a consequence, it is of major importance to optimize the NV$^-$ creation yield for a given nitrogen concentration, while, at the same time, minimizing the creation of other lattice defects.

Room temperature irradiation  can be accompanied by the creation of unwanted paramagnetic defects due to vacancy aggregation  \cite{Irradiation_damage}. In comparison, performing both irradiation and annealing simultaneously - a technique that we will label high temperature (HT) irradiation - provides a more homogeneous procedure in which created vacancies instantaneously become mobile and therefore immediately have the possibility to recombine with nitrogen atoms \cite{hotirr}. In this case, it is expected that the concentration of vacancies will remain continuously low during the full process, which can possibly allow reducing the formation of undesirable defects - e.g. divacancies \cite{Slepetz2014} and vacancy clusters \cite{Slepetz2010}- which induce additional spin decoherence \cite{Favaro2017}. In addition, in the case of bulk crystals, it has been reported a higher NV$^-$ creation after HT irradiation, compared to the case when RT irradiation is applied \cite{hotirr}.
However, an  investigation of possible benefits of  this technique for the production of fluorescent nanodiamonds, in particular in terms of conversion efficiency from P1 to NV$^-$, is missing.

In the present work, we examine the effect of electron  irradiation on  commercially available diamond powder  of different sizes (Microdiamant AG, MSY: $\SI{25}{\nano\meter}$, $\SI{100}{\nano\meter}$,  $\SI{2}{\micro\meter}$), through the use of electron paramagnetic resonance (EPR) and the combination of an atomic force microscope (AFM) with a  confocal microscope.
We first compare, with these techniques,  the RT and HT irradiation in terms of the created quantity of  NV$^-$ and spin properties. We then investigate the NV$^-$ creation yield for  $\SI{100}{\nano\meter}$ and $\SI{2}{\micro\meter}$ particles that underwent different doses of HT irradiation. Systematic analysis of the effect of increasing the irradiation dose is performed, demonstrating   the possibility of reaching a conversion efficiency, defined as the ratio of the final NV$^-$ concentration over the initial P1 concentration, of up to 25\%. Finally,   coherence and relaxation properties of NV$^-$ centers, as well as the creation of additional spin-1 defects,  are discussed.

\section{Results and discussion}

\subsection{Evaluation of the nitrogen content in the starting material}

The samples investigated in this work were obtained from diamond powder produced by the company Microdiamant AG, are of type Ib, and consist of different sizes: \SI{2}{\micro\meter} (Microdiamant, MSY 1.5-2.5), \SI{100}{\nano\meter} (Microdiamant, MSY 0-0.2) and \SI{25}{\nano\meter}  (Microdiamant, MSY 0-0.05). These commercially available  powders are produced through the High Pressure High Temperature (HPHT)  technique, and subsequent milling is performed to obtain smaller size fractions below  \SI{1}{\micro\meter}.

To characterize the effects of irradiation in terms of NV$^-$ formation, it is important to know the amount of nitrogen originally present in the material.  
Continuous wave electron paramagnetic resonance (CW EPR)  spectroscopy has been established as a powerful tool for quantification of paramagnetic defects, such as P1 and NV$^-$ in diamond \cite{Loubser_1978, Shenderova}. In bulk type Ib crystals, this technique allows precise quantification of the nitrogen content, through the detection of P1 signal.
P1 centers are usually quantified and characterized with CW EPR using their characteristic triplet spectrum associated with the hyperfine interaction of the unpaired electron with the $^{14}$N ($I=1$, natural abundance 99.6 \%) nuclear spin. Such a triplet can be conveniently distinguished from other contributions to the spectrum originating, e.g., from spin $1/2$ impurities with different features~\cite{Shames_2015, yavkin2015}.  Following such an analysis (detailed in Supporting Information, section ``P1 Concentration with CW EPR''),  the corresponding spin concentration can be obtained. 
The error on  the determination of P1 spin concentrations with CW EPR was found to be  about $\pm$ 15~\%  for the \SI{2}{\micro\meter} and  $\pm$ 20~\% for the \SI{100}{\nano\meter} and \SI{25}{\nano\meter} samples.
The \SI{2}{\micro\meter} samples used for the present work originate from two different fabrication  batches, and were measured with [P1]$=74 \pm 12$~ppm and [P1]$=53 \pm 8$~ppm respectively, a difference that could potentially be related to variations in the synthesis conditions. The \SI{100}{\nano\meter} and \SI{25}{\nano\meter} samples show respectively $27\pm 5$~ppm and $5.2\pm 1.0$~ppm of P1. Fainting  P1 signal  with  decreasing particle size is consistent with data already reported in the literature for HPHT nanodiamonds obtained upon milling~\cite{yavkin2015,boele2020}. 
One can hypothesize that electron acceptors or donors on the particle surface are responsible for conversion of a fraction of P1 in NDs to nitrogen with a different charge state, or, as it has been proposed, that part of  P1 centers show a different (narrowed) spectrum in NDs  due to   strong exchange interaction with surface dangling bonds~\cite{zegrya2019arxiv}. In the current work, we restrict ourselves to the quantification of the ``core'' P1 centers, that is P1 showing the  characteristic hyperfine pattern due to $^{14}$N interaction. From the values given above, we expect that possible nitrogen that goes undetected with this method cannot play a dominant role in the  NV$^-$ formation mechanism for the  \SI{100}{\nano\meter} and \SI{2}{\micro\meter} particles.  Whether the quantity of created NV$^-$  depends on such nitrogen impurities in smaller, e.g. 25~nm NDs,  would need to be investigated elsewhere.

\subsection{Room versus high temperature irradiation}
\label{section:RT_vs_HT}

To compare the yield of NV$^-$ creation after HT and RT irradiation, we have implemented both methods on diamond powder with particle sizes \SI{25}{\nano\meter}, \SI{100}{\nano\meter} and \SI{2}{\micro\meter}. 
We used in both cases high energy (10 MeV) electron irradiation, which allows to induce a  homogeneous distribution of vacancies over a depth greater than  \SI{1}{\centi\meter}~~\cite{Campbelldamage}, and thus offers the possibility to process large quantities of sample. 
The HT irradiated samples underwent irradiation and simultaneous heating to 800 \textdegree C (see Experimental section). For the RT samples, irradiation at room temperature was followed by annealing at 800 \textdegree C. Samples of a given size received identical electron doses  (\SI{2e18}{\per\square\centi\meter} for the \SI{25}{\nano\meter} sample,  \SI{3e18}{\per\square\centi\meter} for the \SI{100}{\nano\meter} and \SI{2}{\micro\meter} samples).  All samples underwent air oxidation as the final step (see Experimental section).
The NV$^-$ concentration after irradiation was estimated with CW EPR for all samples,  and, for the \SI{25}{\nano\meter} samples, also with a combined AFM-confocal microscope setup (see Experimental section).  We first discuss the CW EPR technique, which allowed systematic characterization of the samples.

NV$^-$ centers, owing to their electron spin $S=1$, can be characterized with EPR. 
It is important to note that the positions of NV$^-$  spectral lines in X-band depend strongly on the diamond orientation, as a consequence of the important zero-field splitting of NV$^-$. Therefore, due to the random orientations of the NDs in powder samples, the EPR spectrum is broadly distributed over a magnetic field range of \SI{2050}{\gauss}  \cite{Shames_2015}. To prevent a too important error on the spin-counting estimate resulting from baseline drift during acquisition, the estimation of NV$^-$ concentration was  performed considering a field region narrower than the full spectrum, as described in section ``Spin-counting with CW EPR'' in the Supporting Information. In the case of the \SI{25}{\nano\meter} samples, the estimation was carried out by considering a $\sim$\SI{50}{\gauss} region centered on one  intensive spectral line, appearing around a field of \SI{2900}{\gauss}. This spectral feature corresponds to NV$^-$ centers oriented perpendicular to the magnetic field, which is the most common case. 
 We found this method of NV$^-$ density determination to have an error of approximately $\pm$20~\%. 
 For the \SI{100}{\nano\meter} and  \SI{2}{\micro\meter} samples, we rather performed double integration on the half-field transition of NV$^-$ \cite{Shames_2015,Shenderova}, which occurs at $\sim$\SI{1650}{\gauss}. Such a method allows reducing further the error on spin-counting, as described in section ``NV$^-$ Concentration with CW EPR'' of the Supporting Information. However,  for these samples, a contribution from other spin-1 defects (non  NV$^-$) appears in the half-field region, which was estimated and subtracted (see section ``Non-NV$^-$ Irradiation-Induced Defects'') to estimate the NV$^-$ concentration. The resulting error for the \SI{2}{\micro\meter} and \SI{100}{\nano\meter} samples was estimated to be $\pm$ 6~\% and $\pm$ 7~\%, respectively.

 \begin{table}[htb!] 
 	\caption{NV$^-$ properties for the \SI{25}{\nano\meter} sample after room and high temperature irradiation.}
 	\begin{adjustwidth}{-.5in}{-.5in}
	 	\begin{center}
	 		\begin{tabular}{ c c c c c c}
	 			\hline
	 			\thead {Irradiation \\ type\\ } & \thead {P1 conc. \\ (ppm)\\ }  & \thead {NV$^-$ conc.,\\ CW EPR \\ (ppm)} & \thead {NV$^-$ conc.,\\ AFM+confocal \\ (ppm)}  &  \thead {$\mathbf{^{\mathrm{NV}}T_2}$ \\ confocal \\ ($\mathbf{\SI{}{\micro\second}})$ }    \\  
	 			\hline
	 			before irradiation & 5.2 & - & - &- \\
	 	    	RT irradiated  &3.5 & 1.0 & 1.6 & 0.5\\ 
	 			\SI{800}{\celsius} irradiated & 1.9 &  2.7 &  2.8 & 0.5 \\ \hline
	 		\end{tabular}
	 	\end{center}
	 	\label{table_RTHT_25nm}
 	\end{adjustwidth}
 \end{table}

  \begin{table}[htb!] 
 	\caption{The comparison of the properties  for the \SI{100}{\nano\meter} sample  irradiated at room and high temperature.}
 	\begin{center}
 		\begin{tabular}{ c c c c c}
 			\hline
 			\thead {Irradiation \\ type} & \thead {P1 conc. \\ (ppm)}  & \thead {NV$^-$ conc. \\ (ppm)}  &  \thead {$\mathbf{^{\mathrm{NV}}T_2}$ \\ ($\mathbf{\SI{}{\micro\second}})$ } &\thead { $\mathbf{^{\mathrm{NV}}T_1}$ \\ ($\mathbf{\SI{}{\milli\second}}$) }   \\  
 			\hline
 			before irradiation & 27 & - & - &- \\
 			RT  irradiated  & 17 & 3.36 & 2.4 & 1.7 \\ 
 			\SI{800}{\celsius} irradiated  & 18 &  3.44 & 2.7& 2.0 \\ \hline
 		\end{tabular}
 	\end{center}
 	\label{table_RTHT_100nm}
 \end{table}
   \begin{table}[htb!] 
 	\caption{The comparison of the properties  for the \SI{2}{\micro\meter} sample  irradiated at room and high temperature.}
 	\begin{center}
 		\begin{tabular}{ c c c c c}
 			\hline
 			\thead {Irradiation \\ type} & \thead {P1 conc. \\ (ppm)}  & \thead {NV$^-$ conc. \\ (ppm)}  &  \thead {$\mathbf{^{\mathrm{NV}}T_2}$ \\ ($\mathbf{\SI{}{\micro\second}})$ } &\thead { $\mathbf{^{\mathrm{NV}}T_1}$ \\ ($\mathbf{\SI{}{\milli\second}}$) }   \\  
 			\hline
 			before irradiation & 74 & - & - &- \\
 		    RT irradiated  & 55 & 6.67 & 2.2 & 2.4 \\ 
 			\SI{800}{\celsius} irradiated   & 50 &  7.41 & 2.2& 2.5 \\ \hline
 		\end{tabular}
 	\end{center}
 	\label{table_RTHT_2um}
 \end{table}
 
 The NV$^-$ concentrations obtained for the different samples (\SI{25}{\nano\meter}, \SI{100}{\nano\meter} and \SI{2}{\micro\meter}) following RT and HT irradiation are given in Tables~\ref{table_RTHT_25nm} to  \ref{table_RTHT_2um}. 
 The case of the \SI{25}{\nano\meter} sample is worth paying attention to, indeed, the  CW EPR results reveal that the quantity of created NV$^-$ is more important in the HT irradiation case, in comparison with RT irradiation.    
To confirm this finding, we implemented an optical quantification method, previously described in Mindarava et al.~\cite{MINDARAVA2020}. This method consists in deducing the number of defects in NDs from the intensity of the photoluminescence (PL) signal in a combined AFM - confocal microscope. This analysis was performed on a set of measurements including 50 single ND particles for each irradiated sample, using \SI{532}{\nano\meter} light for illumination  (see Experimental section for details). To estimate the concentration, the volume of each particle was estimated from their height, measured with the AFM tip, assuming a spherical particle shape. 
Using this method, for the \SI{25}{\nano\meter} sample we found that, on average, RT irradiation created three NV$^-$ centers per particle, while HT irradiation induced  seven NV$^-$ centers per nanodiamond. The corresponding concentrations  match the values obtained in CW EPR (Table \ref{table_RTHT_25nm}). 
We remark that, in comparison with EPR, the optical method suffers from the combined effect of different error sources, including the uncertainty on the particle shape and the strong dependency of NV$^-$ fluorescence on the dipole orientation (the latter leads to fluctuations in the estimated NV$^-$ concentration even after averaging over several particles)~\cite{NVorientation}. Thus CW EPR allows a more precise absolute quantification of NV$^-$ spins than the optical method. In contrast, when considering the \SI{100}{\nano\meter} and \SI{2}{\micro\meter} sizes, we observe no significant difference related to changing the irradiation technique. 
We suggest that our observations could be described by considering the concentration of vacancies in the particles at a given time of  irradiation, and the fact that due to diffusion at high temperature, vacancies might exit the crystal.  Onoda et al.~\cite{onoda2017} determined the  activation energy for vacancy migration   to be 
 $E_m = 2.12$~eV, which  relates to the vacancy diffusion coefficient through the formula $D=D_0 e^{-\frac{E_m}{k_B T}}$  with $D_0=\SI{3.69e-6}{\square\centi\meter\per\second}$. The HT irradiation was performed with a dose rate  \SI{2e13}{\per\square\centi\meter\per\second}. To accumulate a final dose of \SI{2e18}{\per\square\centi\meter}, a total irradiation time $t_i=$\SI{e5}{\second} was needed. At $T=\SI{800}{\celsius}$, the diffusion coefficient is $D_0=\SI{4.07e-16}{\square\centi\meter\per\second}$  
 which yields a root mean square displacement  $l=\sqrt{D t_i} = \SI{63.8}{\nano\meter}$. 
As a consequence, considering the dose applied, a vacancy created at the beginning of the  irradiation process in the \SI{25}{\nano\meter} sample will likely exit the lattice (if it has not recombined meanwhile with a lattice defect).  The probability of that event is significantly lower  in the \SI{100}{\nano\meter} and \SI{2}{\micro\meter} case. Due to the  possibility of the vacancy  exiting the lattice in the \SI{25}{\nano\meter} case, we expect the recombination of two/more vacancies  to form divacancies/vacancy clusters to be less  frequent with HT irradiation, leading to a lower final concentration of such aggregates, as compared to RT irradiation. Such an effect could explain the higher NV$^-$ formation for high temperature irradiation, as   divacancies are responsible for the conversion of NV$^-$ to the neutrally charged NV$^0$~\cite{deak2014}. 
In addition, while the considerations above were made assuming the vacancies have  diffusion properties as in the bulk, the fact that the vacancy migration energy can be lower close to a diamond surface (and thus the diffusion faster) ~\cite{hu2002}, could be an additional explanatory factor for the observed size effect. 

A side effect of the formation of NV$^-$ is the decrease in concentration of P1 spins. From the P1  concentrations given in Tables~\ref{table_RTHT_25nm} to~\ref{table_RTHT_2um} (obtained with CW EPR) one can see that, as expected, the P1 concentration decreases in parallel to the NV$^-$ formation. We expect the creation of one  NV$^-$ to take at maximum two P1 centers, corresponding to the case when one P1 center plays the role  of nitrogen source and another P1 acts as an electron donor:  \mbox{2P1+V$^0$ $\rightarrow$NV$^0$+P1 $\rightarrow$NV$^-$+N$^+$} (V$^0$ is a single neutral vacancy). Remarkably, for the \SI{100}{\nano\meter} and \SI{2}{\micro\meter} samples, it can be seen from Tables~\ref{table_RTHT_100nm} and~\ref{table_RTHT_2um}, that the drop in P1 exceeds  the NV$^-$ concentration concentration by more than a factor of two.  This occurs because NV$^0$ are created as well. Another nitrogen-containing defect, W33, is also formed as we discuss later on  (see section  ``Non-NV$^-$ Irradiation-Induced Defects'').

The $\mathit{^{\mathrm{NV}}T_2}$ times of the \SI{2}{\micro\meter} and \SI{100}{\nano\meter} particles (Tables \ref{table_RTHT_100nm} and \ref{table_RTHT_2um}) were measured with Hahn echo using Pulsed EPR in X-band, on the 2900~G line. For  the \SI{25}{\nano\meter}  powder (Table~\ref{table_RTHT_25nm}), the Hahn echo coherence times were measured at zero magnetic field using a home-built confocal microscope setup (see Experimental section).
The corresponding  $T_2$ times correspond to an average over $8$ different single NV$^-$ centers. The evolution of the $T_2$ times with varying particle size can be described by considering spins both in the bulk (dominantly, P1) and on the surface as possible sources of decoherence.   For \SI{2}{\micro\meter} and  \SI{100}{\nano\meter}, the contribution from P1 needs to be taken into account (see discussion in section  ``NV$^-$ coherence and spin-lattice relaxation times''). 
In the  \SI{25}{\nano\meter}  NDs, decoherence is probably caused \textit{exclusively} by surface spins,  explaining the very short $T_2$ value~\cite{Tisler2009} (for this size, the fact that similar  values are obtained for room temperature and high temperature irradiated samples is probably a consequence of  nearly identical surface states obtained following air oxidation).
For the \SI{100}{\nano\meter} and \SI{2}{\micro\meter}   samples, the \textit{spin-lattice}  relaxation times  $^{\mathrm{NV}}T_1$ were also measured (using Pulsed EPR). We observe a slight shortening of $^{\mathrm{NV}}T_1$ for the \SI{100}{\nano\meter} size in comparison with \SI{2}{\micro\meter}, which could reflect the onset of surface-induced relaxation for NV$^-$~\cite{song2014}.  Overall, the coherence and spin-lattice relaxation times do not show a significant difference between the high and room temperature irradiated samples. The fact that very similar coherence times were observed for the \SI{25}{\nano\meter} samples following RT and HT irradiation suggests that for such small particles, the  surface influence dominates over the effects related to bulk impurities.

Despite the difference between the room and high temperature approaches, only for the \SI{25}{\nano\meter} particles an increase in conversion efficiency upon HT irradiation was found. To explain this effect, we hypothesize that the possibility for the vacancies to diffuse out of the 
particles during irradiation works  in favor of the HT irradiation technique.
In the following sections, we discuss how the  NV$^-$ creation and their spin properties depend on the dose of HT irradiation.

\subsection{Effects of increasing the irradiation dose}
\label{section:vardose}

An important parameter defining the efficiency of NV$^-$ formation is the dose of electron irradiation. Increasing the irradiation dose can increase the conversion efficiency from P1 to NV$^-$. On the other hand, the irradiation dose must be kept under a certain limit as otherwise crystal damage becomes significant~\cite{Irradiation_damage} and the lattice cannot recover its structure even after annealing~\cite{Kalish95}. Therefore, to improve NV$^-$ formation, the irradiation process has to be optimized.

To understand the effect of the irradiation dose on the formation of NV$^-$ defects, we have implemented high temperature irradiation with different doses on  different fractions of the same sample batch.  For this analysis, we selected  \SI{2}{\micro\meter} (MSY2) and \SI{100}{\nano\meter} (MSY0.1) samples. The samples went through the same treatment as the HT irradiated samples described in section \ref{section:RT_vs_HT}, but were subject to varying irradiation doses. The \SI{2}{\micro\meter} samples were irradiated with electron doses of  0.5, 1, 2, 3, 6 and \SI{9e18}{\centi\meter}$^{-2}$, and are named correspondingly 0.5MSY2, 1MSY2, 2MSY2, 3MSY2, 6MSY2, 9MSY2. The \SI{100}{\nano\meter} samples were irradiated with doses of 0.5, 1, and   \SI{3e18}{\centi\meter}$^{-2}$ and are labeled 0.5MSY0.1, 1MSY0.1, 3MSY0.1.

\begin{figure}[h!]
	\centering
	\includegraphics[width=3.25in]{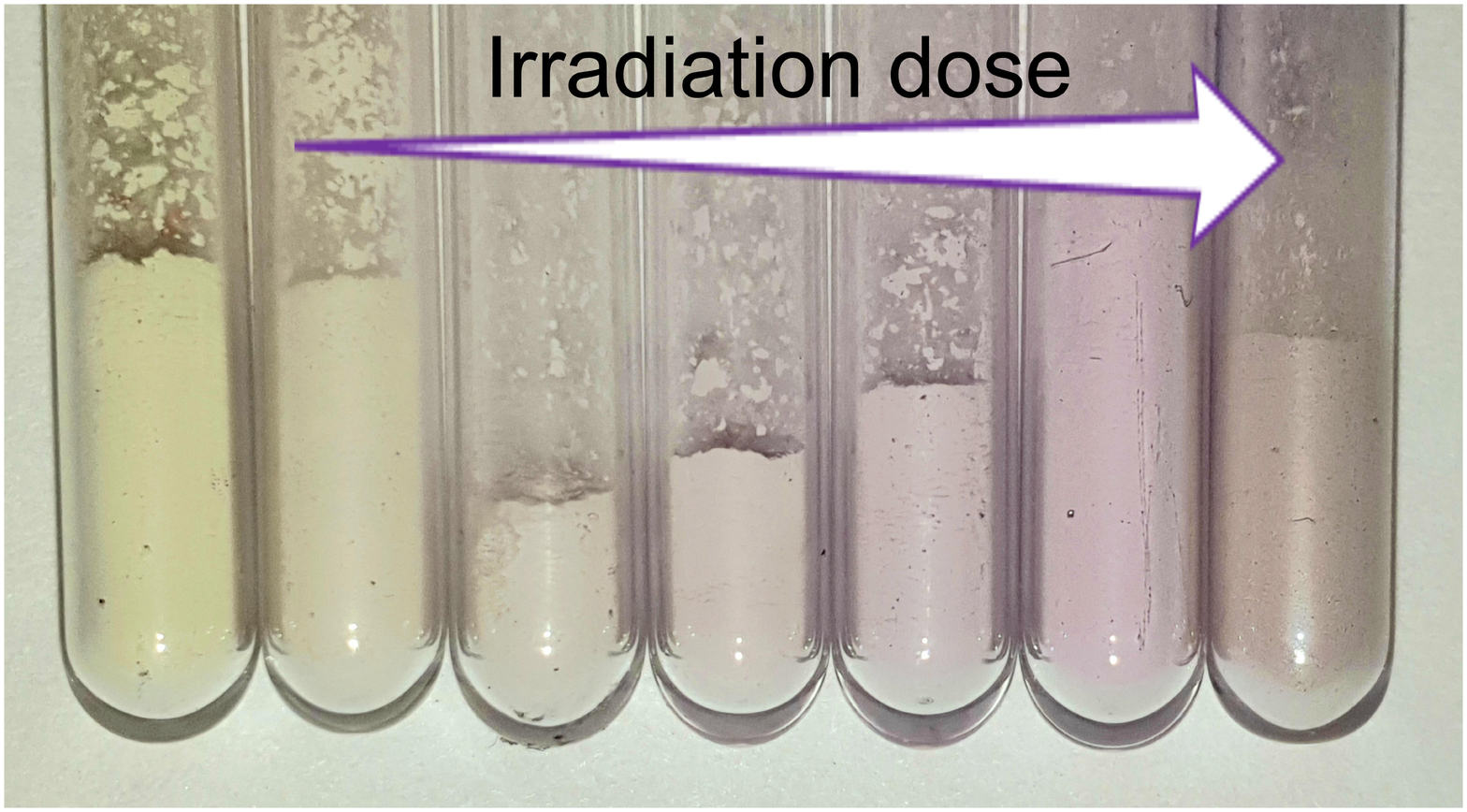}
	\caption[]{A photograph of the samples, HT irradiated with different doses, demonstrating the change in the color of the samples from light yellow to purple with increasing NV$^-$ concentrations. From left to right: the \SI{2}{\micro\meter} sample before electron irradiation, 0.5MSY2 (dose \SI{0.5e18}{\centi\meter}$^{-2}$), 1MSY2 (\SI{1e18}{\centi\meter}$^{-2}$), 2MSY2 (\SI{2e18}{\centi\meter}$^{-2}$), 3MSY2 (\SI{3e18}{\centi\meter}$^{-2}$), 6MSY2 (\SI{6e18}{\centi\meter}$^{-2}$), 9MSY2 (\SI{9e18}{\centi\meter}$^{-2}$).}
	\label{fig:photo}
\end{figure}

A fingerprint of the NV$^-$ concentration is the color of the sample. After cleaning the surface with air oxidation, the non-irradiated samples usually have yellow color, which reveals the presence of P1 centers in the diamond particles. With irradiation, as NV$^-$ centers are created, the color of the sample gradually changes to purple. The continuous alteration in color for the \SI{2}{\micro\meter} powder seen in Figure \ref{fig:photo}  reflects the  increase in NV$^-$ concentration with increasing  irradiation dose. 

\begin{figure}[h!]
	\centering
	\includegraphics[width=5in]{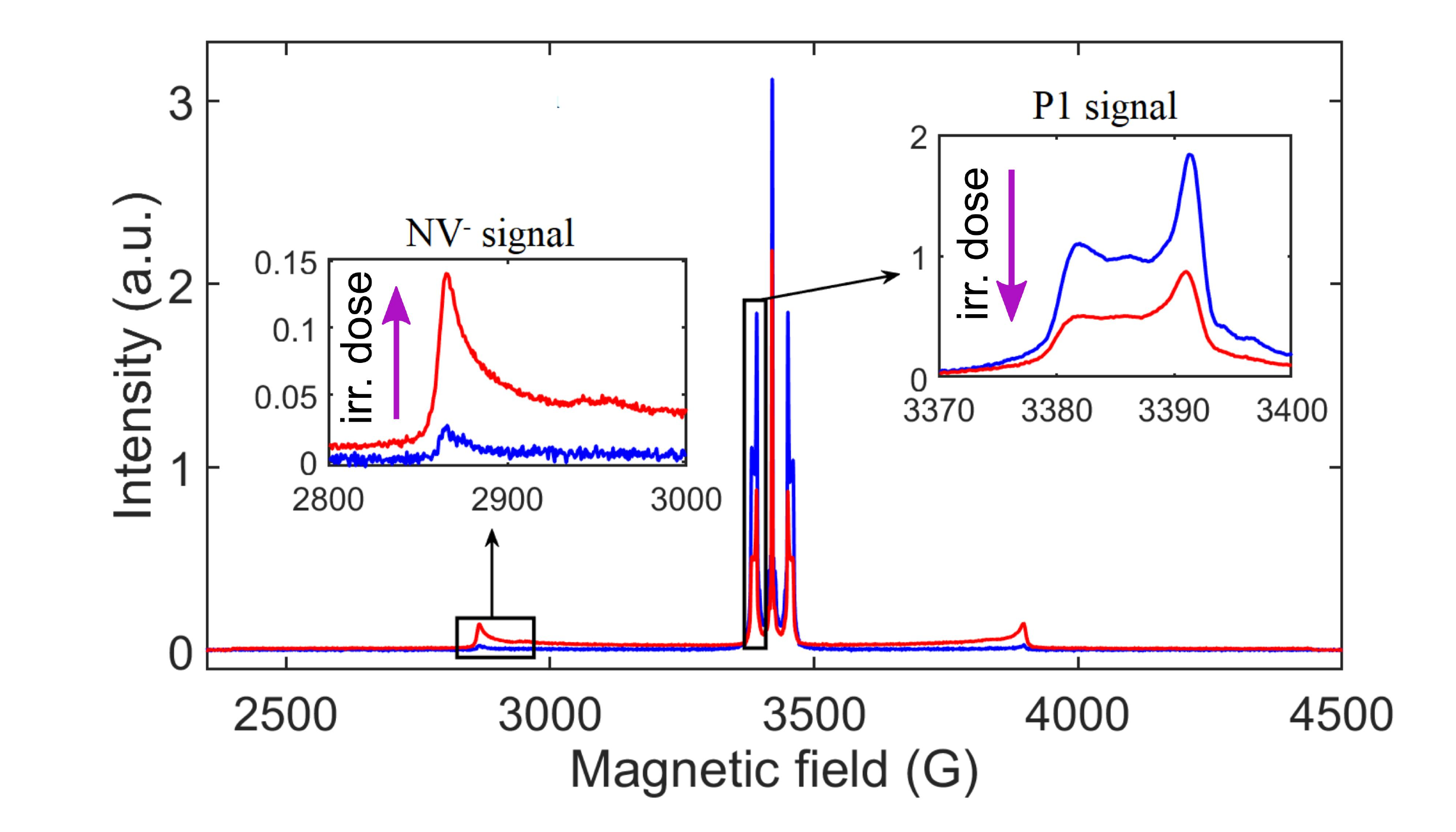}
	\caption[]{NV$^-$ and P1 mass-normalized ESE-EPR spectra for the \SI{2}{\micro\meter} samples with \SI{6e18}{\centi\meter}$^{-2}$ (red curve)  and \SI{0.5e18}{\centi\meter}$^{-2}$ (blue curve) irradiation doses (HT). Insets: low field maximum of NV$^-$ spectra (left) and low field hyperfine line of P1 spectra (right). Change in the respective intensities reflects the conversion from P1 to NV$^-$ induced by irradiation.}
	\label{fig:spec}
\end{figure}

The evolution of P1 and NV$^-$ concentrations has been recorded with CW EPR, which gives the dependencies shown in Tables~\ref{table2} and~\ref{table100nm}. The electron spin echo (ESE) EPR spectra shown in Figure \ref{fig:spec}, measured for two different irradiation doses, illustrate  the efficient conversion from P1 to NV$^-$ induced by irradiation in the case of the  \SI{2}{\micro\meter} powder. The intensity of the NV$^-$ spectrum raises upon increasing the irradiation dose with a simultaneous drop in P1 intensity, suggesting similar evolutions of the respective spin concentrations. 
We remark that the intensities of ESE-detected spectra  are  also affected by the decoherence processes ($T_2$), however, this leads to a stable attenuation factor, depending only weakly on the irradiation dose  (see its evaluation in Supporting Information section ``Pulsed EPR spectrum''). Therefore, the changes in line intensities seen in both insets of Figure \ref{fig:spec} reflect  the conversion from P1 to NV$^-$ occurring with irradiation.

In addition, the P1 concentrations for the \SI{2}{\micro\meter} samples have been estimated using the effect of instantaneous diffusion (ID) with pulsed EPR (see section ``Instantaneous diffusion''  in Supporting Information). The method of ID is based on the dependence of $T_2$ on the flip angle induced by the refocusing pulse. As this method relies on the interaction between close spins \cite{ID1,ID2,Tyryshkin12}, it provides a local  information on the spin density (on the scale of a few nearest neighbors) and is therefore complementary to CW spin-counting. 
 The ID method for P1 centers has been realized on the maximum of the low-field hyperfine pattern in the P1 spectrum. 
To analyze the ID data for P1, as described in the Supporting Information, one needs to take into account the inhomogeneous character of the excitation by the refocussing pulse, which leads to a non-uniform distribution of flip angles in the sample.
The instantaneous diffusion method requires a concentration of the probed species high enough so that the ID effect can compete with other sources of NV$^-$  decoherence~\cite{Tyryshkin12}. 
Therefore, the estimation of NV$^-$ concentration for the samples with a low NV$^-$ density was found not to be possible. For the samples with the highest conversion efficiencies, one can use this technique, as we demonstrated for the 6MSY2 sample (see section ``Analysis of Instantaneous Diffusion for NV$^-$ Centers'' in the Supporting Information), where it gives a result of 10.8 ppm, which is comparable with the result received with CW EPR.

\begin{table}[htb!]
      \centering
	\caption{Summary of P1 and NV$^-$ center concentrations,  and NV$^-$ coherence and relaxation times, for the $\SI{2}{\micro\meter}$ HT irradiated samples with different irradiation doses. The error on  the P1 and NV$^-$ centers  concentrations estimated with CW  EPR is $\pm 15$~\% and $\pm 6$~\%, respectively     (the initial P1 concentration differs from the value seen for the \SI{2}{\micro\meter} sample in Table~\ref{table_RTHT_2um}, which is a consequence of a different fabrication batch being used).  }	
	\begin{adjustwidth}{-.5in}{-.5in} 
		\begin{center}
			\begin{tabular}{  c c c c c c c}
				\hline 
				\thead{Sample\\name}& \thead {Irradiation \\ dose ($\times$\SI{e18}{\centi\meter}$^{-2}$} )  &  \thead  {P1 conc., \\ CW (ppm) } & \thead {P1 conc., \\ ID (ppm)} & \thead {NV$^-$ conc., \\ CW (ppm)}   &  \thead {$\mathbf{^{\mathrm{NV}}T_2}$ \\ ($\mathbf{\SI{}{\micro\second}})$ } &\thead { $\mathbf{^{\mathrm{NV}}T_1}$ \\ ($\mathbf{\SI{}{\milli\second}}$) } \\ 
				\hline
		    	0MSY2 & non-irradiated  & 53  & 64  & -  &   -  & -  \\ 
				0.5MSY2 & 0.5  & 55   & 60 & 1.17 &   2.2 & 2.5  \\ 
				1MSY2 & 1 &   41  & 53 &   2.56 & 2.6 & 2.5 \\
				2MSY2& 2  &   48  & 53 &  3.15 & 2.1 & 2.6\\ 
				3MSY2 & 3  &   40  & 53 &  4.63 &  2.6 &  2.3  \\
				6MSY2 & 6 &   18  & 41 &  10.3 &  2.1  & 2.3  \\   
				9MSY2 & 9 &   13  & 30 &  13.5 &  1.9  & 1.6  \\   \hline	
			\end{tabular}
		\end{center}
	\label{table2}
   \end{adjustwidth}
\end{table}
\begin{table}[htb!]
	\caption{Summary of P1 and NV$^-$ center concentrations, and NV$^-$ coherence and relaxation times, for the $\SI{100}{\nano\meter}$ HT irradiated samples with different irradiation doses. The error on  the determination of the   P1 and NV$^-$  concentration  is $\pm 20$~\% and $\pm 7$~\%, respectively.   }
	\begin{adjustwidth}{-.5in}{-.5in}	
		\begin{center}
			\begin{tabular}{  c c c c c c}
				\hline 
				\thead{Sample\\name}& \thead {Irradiation \\ dose ($\times$\SI{e18}{\centi\meter}$^{-2}$} )  &  \thead  {P1 conc., \\ CW (ppm) } &  \thead {NV$^-$ conc., \\ CW (ppm)}   &  \thead {$\mathbf{^{\mathrm{NV}}T_2}$ \\ ($\mathbf{\SI{}{\micro\second}})$ }
				&  \thead {$\mathbf{^{\mathrm{NV}}T_1}$ \\ ($\mathbf{\SI{}{\milli\second}})$ } \\ 
				\hline
		    	0MSY0.1 & non-irradiated  & 27 & -  & - & -   \\ 
				0.5MSY0.1 & 0.5  & 24   & 0.98 & 3.2 & 2.0  \\ 
				1MSY0.1 & 1 &   22  & 1.97  &  3.1  &   2.2 \\
				3MSY0.1 & 3  &   18  & 3.44 &  2.7 &  2.0  \\
	    \hline	
			\end{tabular}
		\end{center}
		\label{table100nm}
	\end{adjustwidth}
\end{table}

\begin{figure}[h!]
	\centering
	\includegraphics[width=5in]{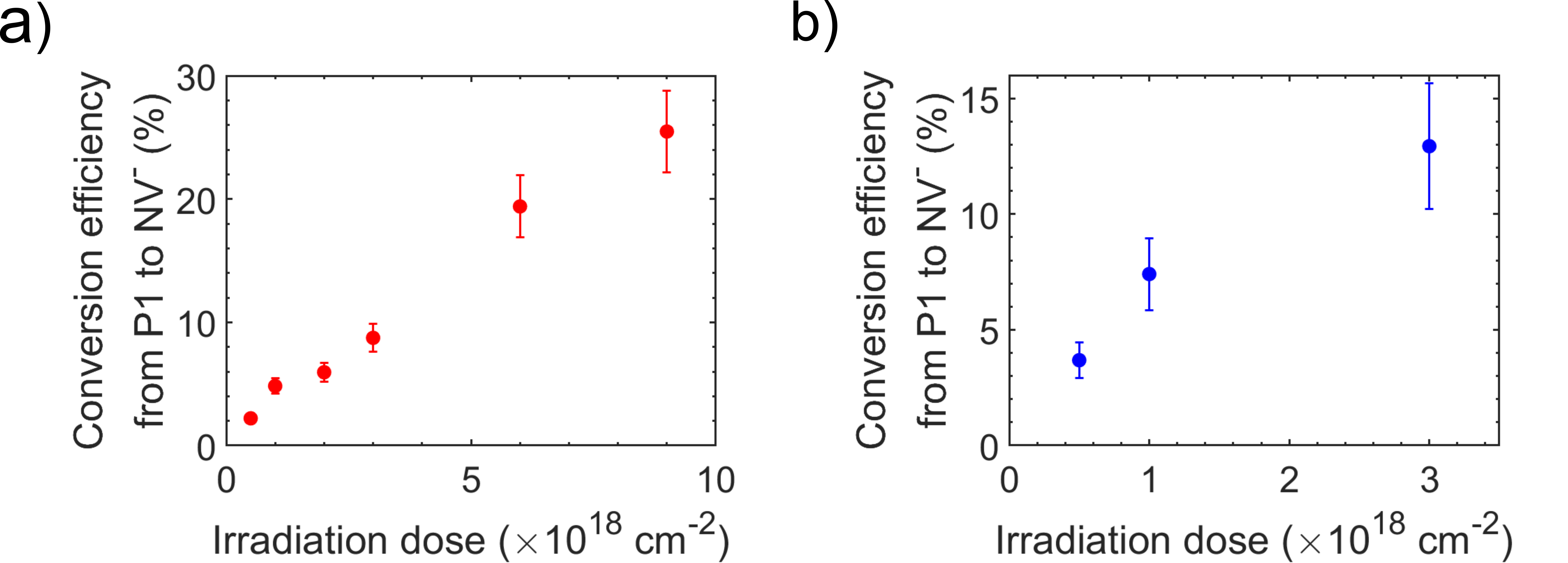}
	\caption[]{Dependence of P1 to NV$^-$ conversion efficiency on the electron irradiation dose with HT irradiation, demonstrating a growth of NV$^-$ creation yield with increasing irradiation dose for the \SI{2}{\micro\meter} (a) and \SI{100}{\nano\meter} (b) samples. The P1 and NV$^-$ concentrations were estimated with CW EPR.}
	\label{fig:conversion}
\end{figure}

The results presented in the Tables \ref{table2} and \ref{table100nm} demonstrate a significant reduction in P1 concentration after irradiation, exceeding the concentration of created NV$^-$ by more than a factor of two. We interpret this as the consequence of several effects, as stated earlier, the creation of one NV$^-$  can take up to two P1 centers and, besides,  other nitrogen-containing defects can be created. In this respect, we could  evidence the creation of NV$^0$, from   the optical photoluminescence spectrum (see section ``Photoluminescence spectra'' in the Supporting Information) and W33 defects  (see following section).

Figure~\ref{fig:conversion} demonstrates the increase in P1 to NV$^-$ conversion efficiency with an increasing electron irradiation dose for the \SI{2}{\micro\meter} and \SI{100}{\nano\meter} diamond powder, reaching values of $25 \pm 3$\% for  \SI{2}{\micro\meter}, at dose  \SI{9e18}{\centi\meter}$^{-2}$. For the for  \SI{100}{\nano\meter} powder, we obtain a conversion efficiency of $13 \pm  2.6$\% , at dose  \SI{3e18}{\centi\meter}$^{-2}$. 
Interestingly, the conversion efficiency for the \SI{2}{\micro\meter} powder at the same dose is $8.7 \pm  1$\%, which is lower.  As the P1 concentration for this sample is higher, our observations are consistent with the trend of decreasing conversion efficiency with increasing P1 content, which was already reported in~\cite{hotirr}.

To increase the NV$^-$ concentration further, one could increase the dose of irradiation.  At high doses, one expects a saturation in NV$^-$  concentration, which we can not clearly observe in Figure~\ref{fig:conversion}.  Further measurements, focusing on fluences above \SI{1e19}{\per\square\centi\meter}, would be needed to determine clearly the saturation parameters, in particular, the maximum attainable NV$^-$  concentration. 
We remark that the application of higher doses  with HT irradiation would imply longer irradiation times and thus longer annealing times. This would increase the  diffusion length of vacancies, which would then have the possibility to diffuse out of the crystal not only for  the  \SI{25}{\nano\meter} size, but also for bigger particles. For instance, with the parameters of our irradiation chamber, a dose \SI{1e19}{\per\square\centi\meter} would  correspond to an annealing time of \SI{5e5}{\second}, which  would result in a diffusion length $l\sim$\SI{140}{\nano\meter} (see section~\ref{section:RT_vs_HT}), meaning a significant  fraction of vacancies would diffuse out of the lattice, considering \SI{100}{\nano\meter} particles.   Following the analysis presented in section~\ref{section:RT_vs_HT}, this could lead to a better conversion efficiency  in comparison with the RT technique, also for this particle size.

In addition, a side effect of high irradiation doses is the creation of non-NV$^-$ paramagnetic defects, as we discuss now.

\subsection{Non-NV$^-$ irradiation-induced defects}
\label{sec:other_defects}

\begin{figure}[h!]
	\centering
	\includegraphics[width=4in]{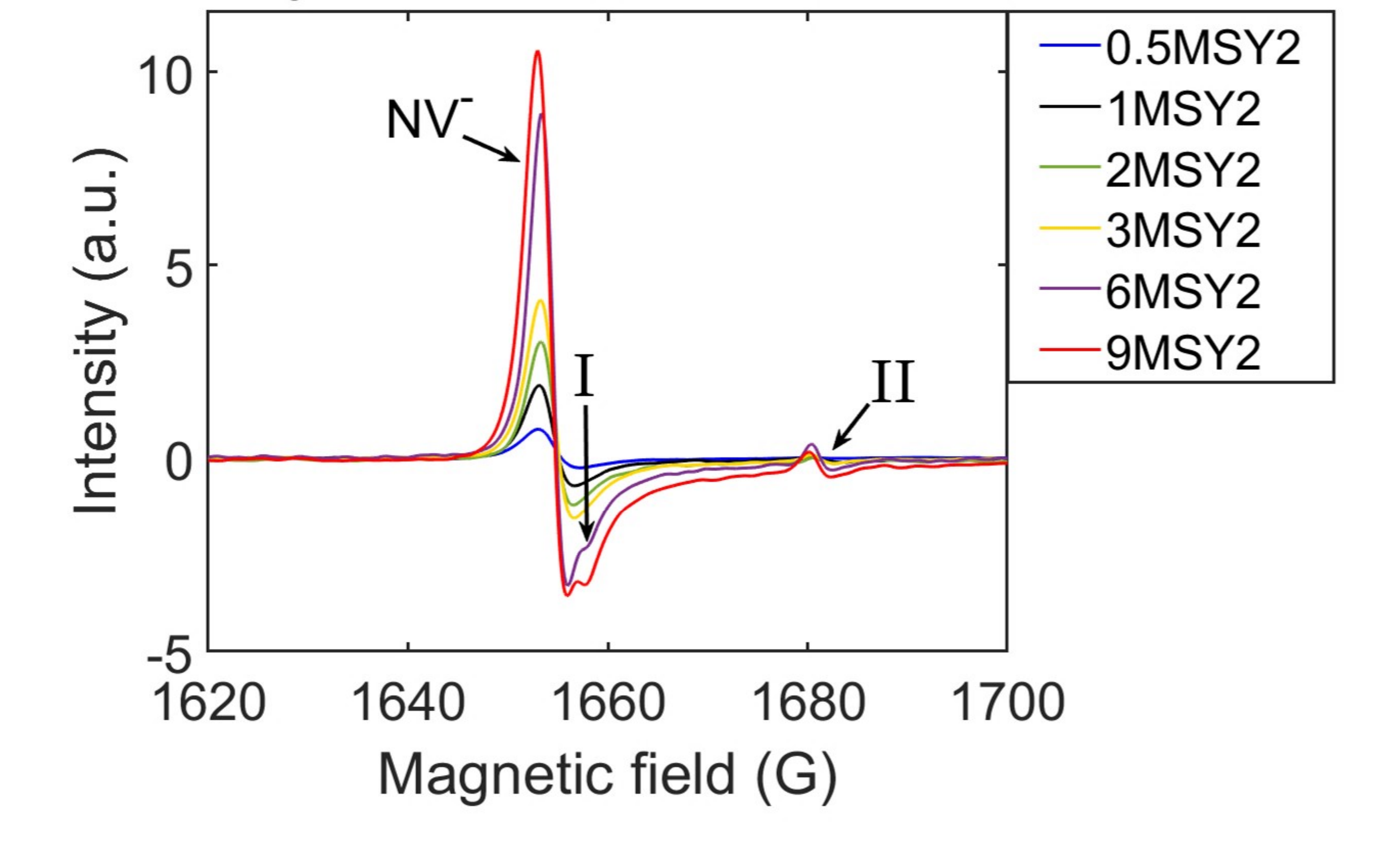}
	\caption[]{Mass-normalized first-derivative CW EPR spectra of the NV$^-$ half-field transition  for the \SI{2}{\micro\meter} HT irradiated samples,  demonstrating a rise of the signal from additional spin-1 defects (\RomanNumeralCaps{1} and \RomanNumeralCaps{2}) with increasing irradiation dose. }
	\label{fig:DQT}
\end{figure}

The formation of paramagnetic defects aside of NV$^-$ is unavoidable during the irradiation process. Here we describe the appearance of such impurities following HT irradiation, for the \SI{2}{\micro\meter} and \SI{100}{\nano\meter} samples discussed in section~\ref{section:vardose}

For these samples, EPR spectra in the half-field region reveal the presence of defects with $S=1$ appearing as additional lines (\RomanNumeralCaps{1} and \RomanNumeralCaps{2}) in the vicinity of the signal from NV$^-$. For the \SI{2}{\micro\meter} samples, Figure \ref{fig:DQT} demonstrates the rise of the intensity of these additional lines, marked as \RomanNumeralCaps{1} and \RomanNumeralCaps{2}, with increasing irradiation dose.
The positions of the lines for the \RomanNumeralCaps{1} and \RomanNumeralCaps{2} defects correspond to the expected positions for W33 and W16 centers, that were  previously reported to appear as a consequence of room temperature irradiation \cite{Shames2017_W33}. 
A fit of the spectrum yields, for the  \SI{2}{\micro\meter} samples, the  concentrations  given in  Table~\ref{table3}. 
In the \SI{100}{\nano\meter} samples, the defect \RomanNumeralCaps{1} (W33) also appears  (Table~\ref{tableW33100nm}),  however, the defect \RomanNumeralCaps{2} (W16) could not be detected.
 The concentrations  were obtained after performing a fit of the CW spectrum including the contributions from NV$^-$, \RomanNumeralCaps{1} and \RomanNumeralCaps{2} defects (see Supporting information, section ``Simulation of the Half-Field Spectrum''), taking the parameters (spin Hamiltonian) for the simulation of W33 and W16 respectively from Nadolinny et al.  \cite{NADOLINNY_W33} and Loubser et al. \cite{Loubser_1978}.
From Tables~\ref{table3} and~\ref{tableW33100nm}, one can see a linear increase of the W16 and W33 concentrations with the irradiation dose, suggesting that the structure of these centers  involve one vacancy, or a split-vacancy with an interstitial atom. The structure of the W33 center in particular has been described by Nadolinny et al. \cite{NADOLINNY_W33} as a nitrogen-vacancy defect with a low symmetry, due to the presence of a positively charged nitrogen (N$^+$) in its vicinity.

\begin{table}[htb!]
	\caption{The concentrations of \RomanNumeralCaps{1}(W33) and \RomanNumeralCaps{2}(W16) defects for the \SI{2}{\micro\meter} HT irradiated samples with different irradiation doses. The \RomanNumeralCaps{2}(W16) defects were not  possible to detect for the 0.5MSY2 sample due to their low concentration.}
	\begin{center}
		\begin{tabular}{c c c c }
			\hline
			\thead{Sample\\ name}& \thead {Irradiation \\ dose ($\times$\SI{e18}{\centi\meter}$^{-2}$)}   &  \thead  {\RomanNumeralCaps{1} center (W33) \\ conc.(ppm) } &  \thead  {\RomanNumeralCaps{2} center (W16)  \\ conc.(ppm) }\\ 
			\hline
			0.5MSY2 & 0.5  &  0.10  & - \\ 
			1MSY2 & 1  &  0.23  &  0.10 \\ 
			2MSY2 & 2  &  0.24  &  0.11 \\ 
			3MSY2 & 3  &  0.42  &  0.15 \\	
			6MSY2 & 6 &  1.14  & 0.57 \\  
		    9MSY2 & 9 &  1.93 &  0.87 \\   		\hline
		\end{tabular}
	\end{center}
	\label{table3}
\end{table}
\begin{table}[htb!]
	\caption{The concentrations of \RomanNumeralCaps{1}(W33) defects for the \SI{100}{\nano\meter} HT irradiated samples with different irradiation doses. For the 0.5MSY0.1 sample,  \RomanNumeralCaps{1}(W33) defects were not possible to detect due to their low concentration.}
	\begin{center}
		\begin{tabular}{c c c}
			\hline
			\thead{Sample\\ name}& \thead {Irradiation \\ dose ($\times$\SI{e18}{\centi\meter}$^{-2}$})  &  \thead  {\RomanNumeralCaps{1} center (W33) \\ conc.(ppm) } \\ 
			\hline
			0.5MSY0.1 & 0.5  &  -  \\ 
			1MSY0.1 & 1  &  0.08  \\ 
			3MSY0.1 & 3  &  0.34 \\	 	\hline
		\end{tabular}
	\end{center}
	\label{tableW33100nm}
\end{table}

\subsection{NV$^-$ coherence and spin-lattice relaxation times}
 
 Estimation of the NV$^-$ coherence and spin-lattice relaxation time ($\mathit{^{\mathrm{NV}}T_2}$ and $\mathit{^{\mathrm{NV}}T_1}$) for the $\SI{2}{\micro\meter}$ and \SI{100}{\nano\meter} samples are given in Table~\ref{table2} and \ref{table100nm}, respectively.

\begin{figure}[h!]
	\centering
	\includegraphics[width=5.5in]{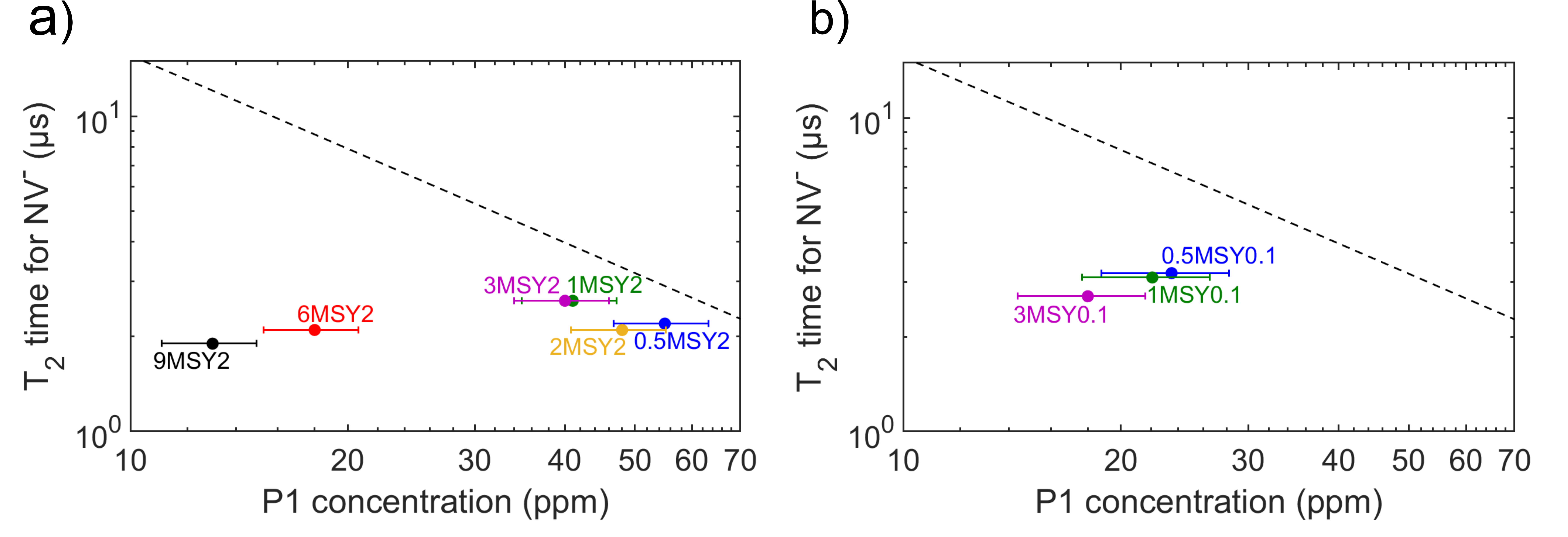}
	\caption[]{$\mathit{^{\mathrm{NV}}T_2}$ times obtained for the  $\SI{2}{\micro\meter}$ (a) and $\SI{100}{\nano\meter}$ (b) HT  irradiated samples, as a function of P1 concentration, compared to the prediction  from Bauch et al.~\cite{bauch2019} for  $\mathit{^{\mathrm{NV}}T_2}$ in case the NV$^-$-P1 interaction is the dominant  decoherence factor (dashed line on both plots). The deviation of the  data points  from the dashed line demonstrates the presence of additional sources for NV$^-$ decoherence besides P1 centers.}
	\label{fig:T2_vs_P1}
\end{figure}

One important source of NV$^-$ decoherence in  non-irradiated bulk material or microdiamonds is the interaction with P1 centers~\cite{bauch2019}, which can be  supplanted by sources in the vicinity of the surface (e.g. dangling bonds) for small NDs~\cite{Tisler2009, tsukahara2019}. After irradiation, however, NV$^-$ centers and potentially other  irradiation-induced defects (see previous section) can themselves be a source of  decoherence. To distinguish the contributions  from P1 centers and from other sources of NV$^-$ decoherence, we used the formula from Bauch et al.~\cite{bauch2019}, that corresponds to a fit of the dependence of $^{\mathrm{NV}}T_2$ on the P1 concentration for single crystalline diamond with low NV$^-$ content. This dependence reads $^{\mathrm{NV}}T_2 = C/\textrm{[P1]}$, as a function of P1 concentration, with $C=160\pm12\ \SI{}{\micro\second}\cdot\textrm{ppm}$~\cite{bauch2019}. From  Fig.~\ref{fig:T2_vs_P1}, we can draw several conclusions.  
First, looking at the samples with the lowest irradiation doses for particle sizes  \SI{2}{\micro\meter} and  \SI{100}{\nano\meter} (0.5MSY2 and 0.5MSY0.1), we observe that the discrepancy to the prediction is higher for the  \SI{100}{\nano\meter} sample. We interpret this as a sign that the surface already plays a role in decoherence for \SI{100}{\nano\meter} particles, but has negligible effect in the \SI{2}{\micro\meter} case. We emphasize that such conclusion could only be made after simultaneously considering $T_2$ and P1 concentration, the latter being different between the  \SI{2}{\micro\meter} and \SI{100}{\nano\meter} samples.
Second, as the irradiation dose is increased, the measured data  drifts away from the prediction, as the decrease in P1 concentration does not translate into a prolonged $T_2$ time. This demonstrates the growing influence of ``non-P1'' sources  appearing as the irradiation dose is increased.  Among these sources are NV$^-$ themselves, acting both through spectral and instantaneous diffusion~\cite{Tyryshkin12}. Effects of NV$^-$-NV$^-$ interactions are naturally expected to become more important with increasing irradiation doses. As a confirmation of their role in decoherence, we could establish, for the 6MSY2 sample, that removing the effect of instantaneous diffusion between NV$^-$ yields an  increase in $T_2$ from \SI{2.1}{\micro\second} to \SI{3.8}{\micro\second}  (see section ``Analysis of Instantaneous Diffusion for NV$^-$ Centers'' in  Supporting Information), which accounts for part of the deviation to the dashed line in Fig.~\ref{fig:T2_vs_P1}a. Additionally, spectral diffusion, induced by NV$^-$ or by other defects such as the W16 and W33 center (see section ``Non-NV$^-$ irradiation-induced defects''), can also play a role. Last, in the \SI{100}{\nano\meter} case, for which the $T_2$ seems to \textit{shorten} as a result of irradiation, already at the lowest doses (which is not observed in the \SI{2}{\micro\meter} case), one cannot exclude the creation of new dangling bonds towards the surface. 
Further work would be needed to distinguish the respective contributions to decoherence, for the different particle sizes.

The NV$^-$ spin-lattice relaxation time, 
$^{\mathrm{NV}}T_1$, measured for the  $\SI{2}{\micro\meter}$ samples (Table~\ref{table2}), may yield information on irradiation damage. We observe a drop in $^{\mathrm{NV}}T_1$  past the  irradiation dose  \SI{6e18}{\centi\meter}$^{-2}$ (from 2.3 to 1.6~ms). Because of the particle size, we suppose any influence from surface defects can be excluded. As a hypothesis, the creation  of fastly relaxing ($T_2 \sim 1$~ns)  paramagnetic impurities in the crystal under irradiation could explain the shortening of $\mathit{^{\mathrm{NV}}T_1}$  (owing to their very short $T_2$ time, such impurities are difficult to detect directly with EPR).  Further studies performed with higher irradiation doses would be needed to confirm this trend.

\section{Experimental section}

\subsection{Sample preparation}

Commercially available micro- and nanodiamonds with a mean particle size of  $\SI{2}{\micro\meter}$ (Microdiamant, MSY 1.5-2.5, type Ib, HPHT), $\SI{100}{\nano\meter}$ (Microdiamant, MSY 0-0.2, type Ib, HPHT)  and  $\SI{25}{\nano\meter}$ (Microdiamant, MSY 0-0.05, type Ib, HPHT) have been used with size distributions between 1.5-$\SI{2.5}{\micro\meter}$, 0-$\SI{200}{\nano\meter}$ and 0-$\SI{50}{\nano\meter}$, respectively.
For the RT irradiated sample a 10 MeV electron accelerator (MB10-30MP-Mevex Corp., Stittsville Canada) operating under air atmosphere was used, which includes a  permanent cooling to regulate the treatment temperature below 300 \textdegree C. After irradiation the sample was annealed in argon atmosphere at 800 \textdegree C for 5 hours. 

The HT irradiation of the samples has been implemented within a ceramic holder placed in a quartz furnace under permanent argon flow. During warm up of the linear accelerator (also producing 10~MeV electrons) the quartz tube is flushed for approximately 30 minutes with argon. During irradiation, the argon flow is maintained at about 150 ml/min by means of a flow controller (GFC171 from Analyt), keeping the argon pressure close to \SI{1}{\Bar}. The treatment temperature was regulated by the dose per pulse and repetition frequency of the accelerator and was monitored by applying a thermocouple connected to the sample holder and regulated to be 800 \textdegree C. The HT irradiation dose rate is about \SI{2e13}{\centi\meter}$^{-2}$ {s}$^{-1}$.

Surface graphitization occured, either during HT irradiation or, in the case of the RT irradiated sample, during  annealing. Therefore, after  irradiation,  all samples were subjected to air oxidation at 620 \textdegree C for 5 hours to remove the graphitic residues from the surface.

\subsection{Optical measurements} 

\indent For optical and AFM investigation of the $\SI{25}{\nano\meter}$ ND samples (with HT and RT irradiation) $\SI{5}{\micro\liter}$ of ND solution in demineralized water were spin-coated (5000 rpm for 40 seconds) on a plasma cleaned glass substrate. Microwaves were applied on NV$^-$ centers through a \SI{20}{\micro\meter} thick copper wire.\\
\indent Fluorescence and NV$^-$ spin properties were measured with a home-built confocal microscope, where a \SI{532}{\nano\meter} laser was used for NV$^-$ excitation. The setup was controlled using the Qudi software package \cite{Qudi}. The fluorescence was collected through an oil-immersion objective (Olympus UPlanSApo 60x oil NA=1.35). After the bandpass filter with transmission between \SI{625.5}{\nano\meter} and \SI{792.5}{\nano\meter}  the fluorescence was detected with an avalanche photodiode with single photon resolution (Excelitas Technologies). The pulse sequence for the $T_2$ time measurement was implemented using an Arbitrary Waveform Generator (Tektronix AWG70001A).

\subsection{EPR measurements}
   
\indent X-band (9.6 GHz) CW and Pulse EPR  measurements were implemented at room temperature on a Bruker Elexsys E580 EPR spectrometer with waveguide resonator (ER-4122MD4) and FlexLine resonator (ER-4118X-MD5), respectively. The spin-counting in CW has been performed with Bruker software (xEPR). 
Low microwave power was used in order to avoid saturation of the detected signal: <$\SI{3}{\micro\watt}$ for measuring P1 and the ``allowed'' transition of NV$^-$ (at 2900 G),   and <$\SI{20}{\micro\watt}$ for the half-field transition of NV$^-$. Experiments were performed with a decoupled cavity ($Q= 8000-10000$).

The samples were measured inside of a quartz EPR tube from Wilmad-Labglass (707-SQ-100M) with an inner diameter of $\SI{3}{\milli\meter}$. Simulations of EPR spectra were performed using the EasySpin Matlab toolbox \cite{Stoll06}. 

Details on the signal acquisition procedures and on the spectral simulation  parameters are provided in Supporting Information.

\section{Conclusions}
A high NV$^-$ density is crucial for potential applications of NDs as fluorescent markers, to increase magnetometry sensitivity and for efficiency of techniques for $^{13}$C nuclear spin hyperpolarization. The results presented here demonstrate the possibility to reach a high formation yield  of NV$^-$ defects in nano- and microdiamonds by implementing simultaneous electron irradiation and annealing.  
 As we hypothesize, the  possibility of the vacancies to diffuse out of the crystal during this `high temperature irradiation''  process  explains the higher NV$^-$ formation observed in the case of  \SI{25}{\nano\meter} nanodiamonds. 
For bigger (\SI{2}{\micro\meter}) particles, we demonstrated that a conversion efficiency of P1 to NV$^-$ of 25 \% can be achieved, a figure that could potentially be increased by using higher irradiation doses. We observed, in \SI{100}{\nano\meter} and  \SI{2}{\micro\meter} samples, the concomitant appearance of additional irradiation-induced spin-1 defects involving one vacancy, identified as W16 and W33 centers.
Despite the creation of such defects, long NV$^-$ coherence and spin-lattice relaxation times prove that no severe irradiation damage has been caused. 
We expect that the presented irradiation technique will allow synthesis of fluorescent nanodiamonds with tailored NV$^-$ concentration, providing opportunity for their applications in nanoscale optical imaging, as magnetic sensors, and for nuclear spin hyperpolarization.

\section*{Acknowledgements}

This work was supported by the DFG (CRC 1279), EU HYPERDIAMOND (Project ID 667192), VW Stiftung (No. 93432), BW Stiftung (No. BWINTSF\RomanNumeralCaps{3}-042), BMBF (Project No. 13N14438, 13GW0281C, 13N14808, 16KIS0832, 13N14810, 13N14990), ERC (Grant No. 319130), JSPS-KAKENHI (No. 17H02751).
We thank Dr. Yan Liu for providing the confocal setup for optical measurements.

%
%

\bibliography{YM_HT_irradiation_Carbon}


\end{document}


\title{{\huge Supporting Information}\\ {\Large Efficient Conversion of Nitrogen to Nitrogen-Vacancy Centers in Diamond Particles with High-Temperature Electron Irradiation}}

\author[a]{Yuliya Mindarava\corref{mycorrespondingauthor}}
\cortext[mycorrespondingauthor]{Corresponding author. Tel: +4915203294898}
\ead{yuliya.mindarava@uni-ulm.de}
\author[a]{ R\'{e}mi Blinder\corref{mycorrespondingauthor1}}
\cortext[mycorrespondingauthor1]{Corresponding author. Tel: +497315023728}
\ead{remi.blinder@uni-ulm.de}
\author[b]{ Christian Laube}
\author[b]{ Wolfgang Knolle}
\author[b]{Bernd Abel}
\author[c]{Christian Jentgens}
\author[d]{Junichi Isoya}
\author[e]{Jochen Scheuer}
\author[a]{Johannes Lang}
\author[e]{Ilai Schwartz}
\author[f]{Boris Naydenov}
\author[a,g]{Fedor Jelezko}
\address[a]{Institute for Quantum Optics, Ulm University, Ulm 89081, Germany}

\address[b]{Department of Functional Surfaces, Leibniz Institute of Surface Engineering, Leipzig 04318, Germany}
\address[c]{Microdiamant AG, Kreuzlingerstrasse 1, CH-8574 Lengwil, Switzerland}
\address[d]{Faculty of Pure and Applied Sciences, University of Tsukuba, Tsukuba, 305-8573, Japan}
\address[e]{NVision Imaging Technologies GmbH, Ulm 89081, Germany}
\address[f]{Institute for Nanospectroscopy, Helmholtz-Zentrum Berlin f\"{u}r Materialen und Energie GmbH, Berlin 14109, Germany}
\address[g]{Centre for Integrated Quantum Science and Technology (IQST), Ulm 89081, Germany}

\maketitle

\section{Spin-counting with CW EPR}

In CW EPR, one employs phase-sensitive detection tuned to the frequency of the magnetic field modulation, which leads to recording the first derivative of the absorption signal. The most direct spin quantification method involves performing double integration on a spectrum acquired in the so called ``low microwave power'' regime. Indeed, when measuring at low microwave power $P$, that is in absence of saturation effects, the signal response is known to be proportional to the total number of spins in the sample and to $\sqrt{P}$ \cite{eaton_2010book}. 
 Different choices for the double integration region however result in different errors on the outcome of spin-counting, as we describe in this section.

\subsection{Sources of error in spin-couting}
\label{SI:CW_error}

Two important sources of error in CW quantitative EPR are \cite{eaton_2010book}: 
\begin{enumerate}
\item the drift of the baseline, that occurs for instance as a consequence of room temperature changes or air drafts around the cavity during the measurement  (a single measurement took from one minute to a few hours),
\item the uncertainty on the determination of the cavity $Q$ value.
\end{enumerate}

To decrease the error resulting from 1., one typically performs subtraction of a properly defined baseline. In addition to taking into account the changes in the cavity response or environment, baseline subtraction can allow removing a background signal originating from spin species,  that should not be included in the spin-counting (such as non-P1 spin $1/2$ defects, see section \ref{SI:CW_P1}). 

Baseline definition was found to be possible, except under the following conditions. First, when the measured spectrum spans over a broad field range. In that case,  the time necessary to perform one sweep over the full spectrum is long, as a result, complex and unpredictable baseline distorsions (which can hardly be subtracted)  can occur. Second, when, from knowledge of the measured spin species, it is predicted that the signal do not decay to zero towards the boundaries of the measured field range.  
When both conditions defined above were avoided (that is when measuring over a narrow field range in a relatively fast time, with the predicted EPR signal decaying to zero on both sides within the range boundaries), the baseline could be defined by interpolation from the two outermost regions of the spectrum using a polynomial function (linear, or quadratic), and subsequently subtracted. A residual dispersion of the double integral values  on distinct measurements \textit{after subtraction}  could still be found. We attribute this dispersion to imperfections in the baseline definition (e.g., non-linear or non-quadratic baseline). 

The $Q$ factor was determined using the readout in the tuning panel, within the  xEPR software (setting attenuation at 33 dB), and was typically in the range $Q=8000-10000$.  The error on the $Q$ factor determination for a given experiment was determined to be $\pm 6$\%.  In the low power regime, the EPR signal is directly proportional to $Q$~\cite{eaton_2010book}, therefore this yields directly a $\pm 6$\% error on the spin-counting, which should be considered in addition to the baseline-related error. Double integration was performed under the Spin-counting tool in xEPR.

\subsection{P1 Concentration with CW EPR}
\label{SI:CW_P1}

The P1 spectrum consists of group of three lines, as a result of the hyperfine coupling to the $^{14}$N nuclear spin ($I$=1), appearing in the $g\approx 2$ region (at X-band frequencies, $f_X=9.4-9.8$ GHz, $g=2$ corresponds to 3350-3500 G). 

The P1 spectrum overlaps with contributions originating for other spin 1/2 species, so that the full spin 1/2 spectrum can be described as the superposition of P1 spectrum with two additional spin $1/2$ components of different widths \cite{yavkin2015, boele2020}. 
First, a broad component that was attributed to deformation-induced defects at the vicinity of the surface~\cite{yavkin2015}. The weight of this  contribution  becomes naturally more important with decreasing particle size. In addition to that, a narrow contribution exists, which could correspond to a P1 spectrum narrowed by interaction with surface electrons~\cite{zegrya2019arxiv}. 

For \SI{2}{\micro\meter} and \SI{100}{\nano\meter} particles, the overlap of these additional components remains weak  in the region of the P1 \textit{hyperfine} lines ($m_I=\pm 1$). The low field hyperfine line was considered for spin-counting. The contribution from other spin $1/2$ defects was approximated in this region as varying linearly with field (on the 1$^{\textrm{rst}}$ derivative spectrum), and was removed as a part of the baseline subtraction procedure, implemented by default in the `Spin-counting' tool in xEPR  (the baseline was defined by interpolation from regions located on each side on the P1 hyperfine line).  The obtained concentration was multiplied by three to account for the three possible states of the $^{14}$N nuclear spin. Including uncertainties from  both   the baseline and $Q$ factor determination, the error was determined to be about $\pm 15$~\% for \SI{2}{\micro\meter} samples and   $\pm 20$~\% for \SI{100}{\nano\meter} samples. 

For \SI{25}{\nano\meter} particles, the background originating from non-P1 defects is dominant.  
The double integral was taken on the full spectrum, P1 and other spin $1/2$ defects. The P1 contribution to the double integral was extracted from a fit of the spectrum performed with the  EasySpin package in Matlab \cite{EasySpin}, including P1 and two other spin 1/2 components.  The resulting  error for P1 density estimation is estimated to be $\pm 20$~\%, coming dominantly from the fit error.



\subsection{NV$^-$ Concentration with CW EPR}

\begin{figure}[!h]
	\centering
	\includegraphics[width=5in]{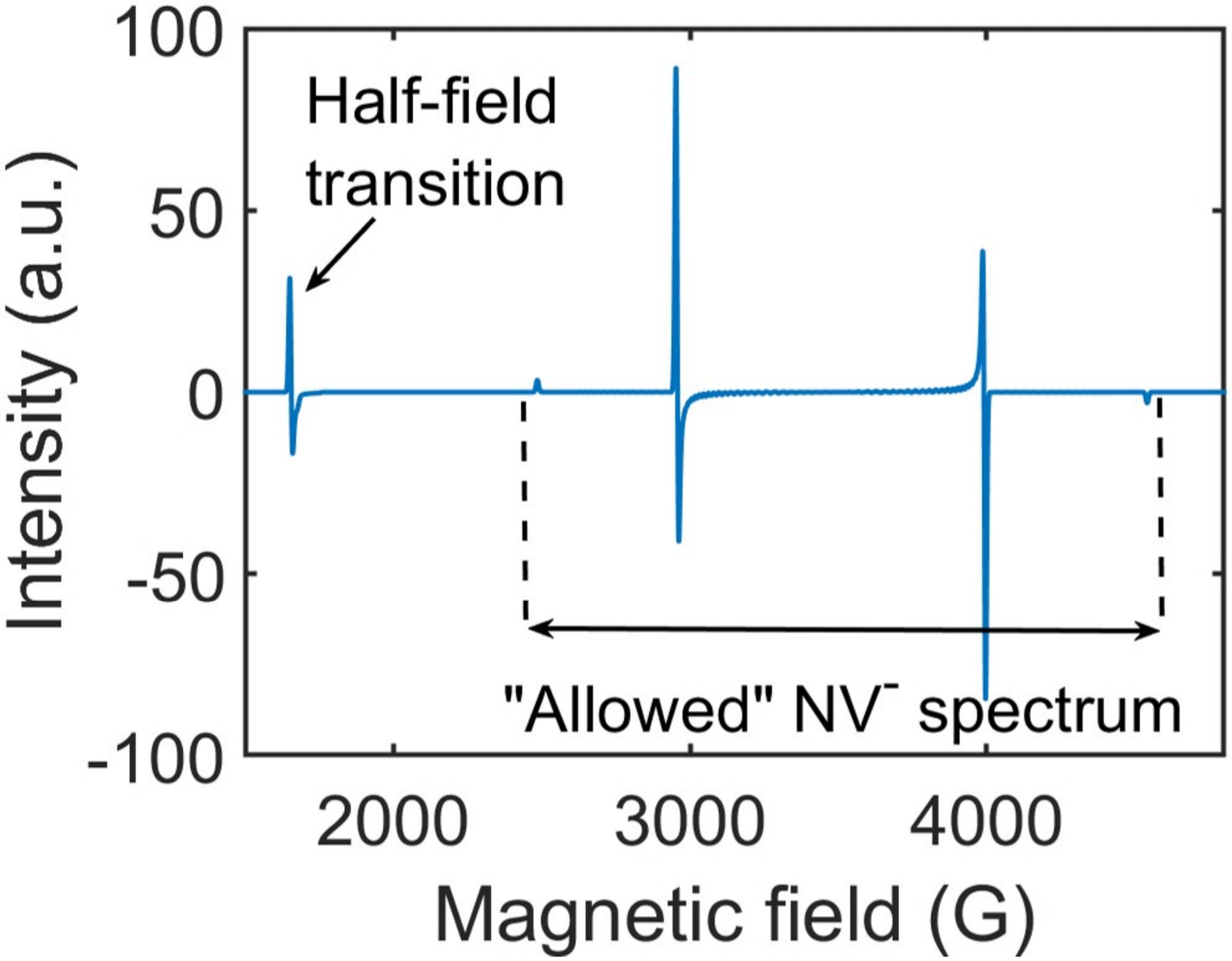}
	\caption{Simulation of the full X-band (9.6 GHz) EPR NV$^-$ spectrum.}
	\label{fig:Int-part}
\end{figure}

The powder spectrum  of NV$^-$ is very broad in X-band, it spans over 2050~G  only for the ``allowed'' transitions, as shown in Fig.~\ref{fig:Int-part}. This complicates the concentration determination with CW EPR, since the baseline could not be defined over a wide region without important error.  This can be circumvented by choosing an integration region that covers only part of the spectrum, as we now discuss.

For the 25 nm nanodiamonds, the NV$^-$ concentrations have been estimated by  integration of the low-field most intensive part in the NV$^-$ spectrum (about 2900 G). The total NV$^-$ concentration was evaluated by multiplying the received result by a factor estimated from the simulation of NV$^-$ spectrum (Figure \ref{fig:Int-part}). However, this method of NV$^-$ density evaluation has still a significant error coming from the baseline drift in the chosen spectral region. Indeed, since the EPR signal is non-zero at the lower and upper-field end of the measured spectral range, the conditions for baseline definition (defined in  section \ref{SI:CW_error}) are not met. In that case the baseline was not subtracted.  The resulting error was estimated to be $\pm 20~\%$.

In contrast, the NV$^-$ spin counting for the $\mathbf{\SI{100}{\nano\meter}}$ and  $\mathbf{\SI{2}{\micro\meter}}$ samples was carried out by double integration on  the \textit{half-field} spectrum of NV$^-$ (around 1660 G)~\cite{Shames_2015,Shenderova}.  As the half-field lines  occur within a narrower field range in comparison with the full ``allowed'' spectrum (Figure \ref{fig:Int-part}), the baseline definition  method presented in section \ref{SI:CW_error} could be applied.  

In the integration region,  the  spectrum consists, however, of contributions both from NV$^-$ and from other irradiation-induced spin-1 defects. To determine the weight of the NV$^-$ contribution we have performed the fit of the experimental spectrum with a simulation including both NV$^-$ and the other irradiation-induced defects (see section ``Simulation of the half-field spectrum''). 
Finally, to estimate the full NV$^-$ spin concentration, as with the previous method, one must take into account a  multiplication factor. Here it corresponds to the ratio of the double integrals over ``allowed'' and half-field transitions, respectively. That factor was calculated from a simulation of the NV$^-$ spectrum (see Figure \ref{fig:Int-part}), which yields a value of 41. The error on the concentration was estimated to be $\pm 6~\%$ for the $\mathbf{\SI{2}{\micro\meter}}$ samples and $\pm 7~\%$ for the $\mathbf{\SI{100}{\nano\meter}}$ samples, coming dominantly from the uncertainty on the determination of the EPR cavity $Q$-value.

\section{Simulation of the Half-Field Spectrum}
\begin{figure}[h!]
	\centering
	\includegraphics[width=\linewidth]{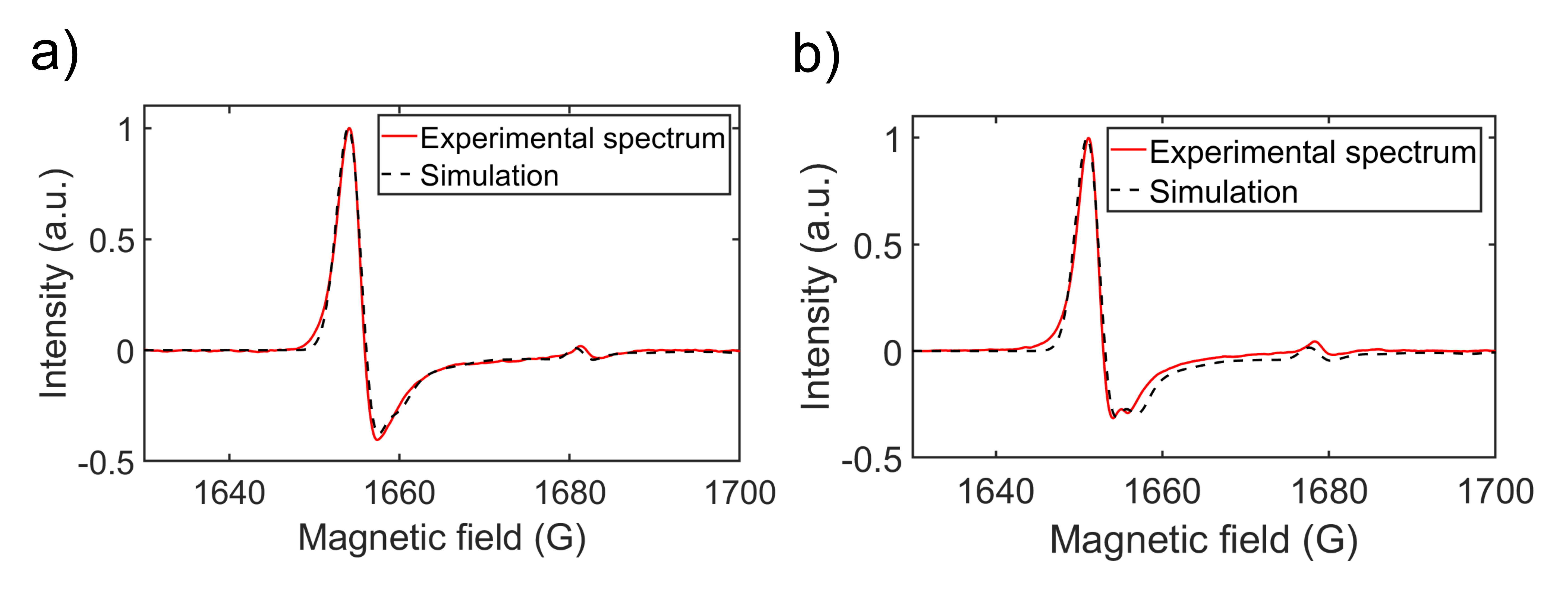}
	\caption{The experimental and simulated spectra of the half-field transition lines of NV$^-$ and irradiation-induced defects \RomanNumeralCaps{1} (W33) and \RomanNumeralCaps{2} (W16) for the 9MSY2 (a) and 3MSY2 (b) samples.}
	\label{fig:Simul_irr_defects}
\end{figure}

To estimate the concentrations of NV$^-$ (see Table \ref{table2} ), and of the \RomanNumeralCaps{1} (W33) and \RomanNumeralCaps{2} (W16) defects (see Tables \ref{table3} and \ref{tableW33100nm}) we fitted the spectra close to half-field transition of NV$^-$ with the simulation composed of the simulated spectra for NV$^-$, W33 and W16 defects (see Figure \ref{fig:Simul_irr_defects}), where the parameters for the last two centers were taken from Nadolinny et al. \cite{NADOLINNY_W33} and Loubser et al. \cite{Loubser_1978} (see Table \ref{tableS2}). The simulations have been  performed with the EasySpin package in Matlab \cite{EasySpin}. 

\begin{table}[htb!]
	\caption{The parameters for the simulation of spectra for the irradiation-induced defects and NV$^-$ centers}	
	\begin{center}
		\begin{tabular}{ c c c c }
			\hline 
			Center & g-value &  D (MHz)  & E (MHz) \\ 
			\hline
			\RomanNumeralCaps{1} (W33)  & 2.0028  & 2792.4 & 247.0 \\ 
			\RomanNumeralCaps{2} (W16)  & 2.0026 & 2490.1 & 16.2\\ 	
			NV$^-$  & 2.0028 & 2874.0 & -\\ 
			\hline
		\end{tabular}
	\end{center}
	\label{tableS2}
\end{table}

\clearpage
\section{$T_2$ and $T_1$ measurements}

\begin{figure}[htb!]
	\centering
	\includegraphics[width=3in]{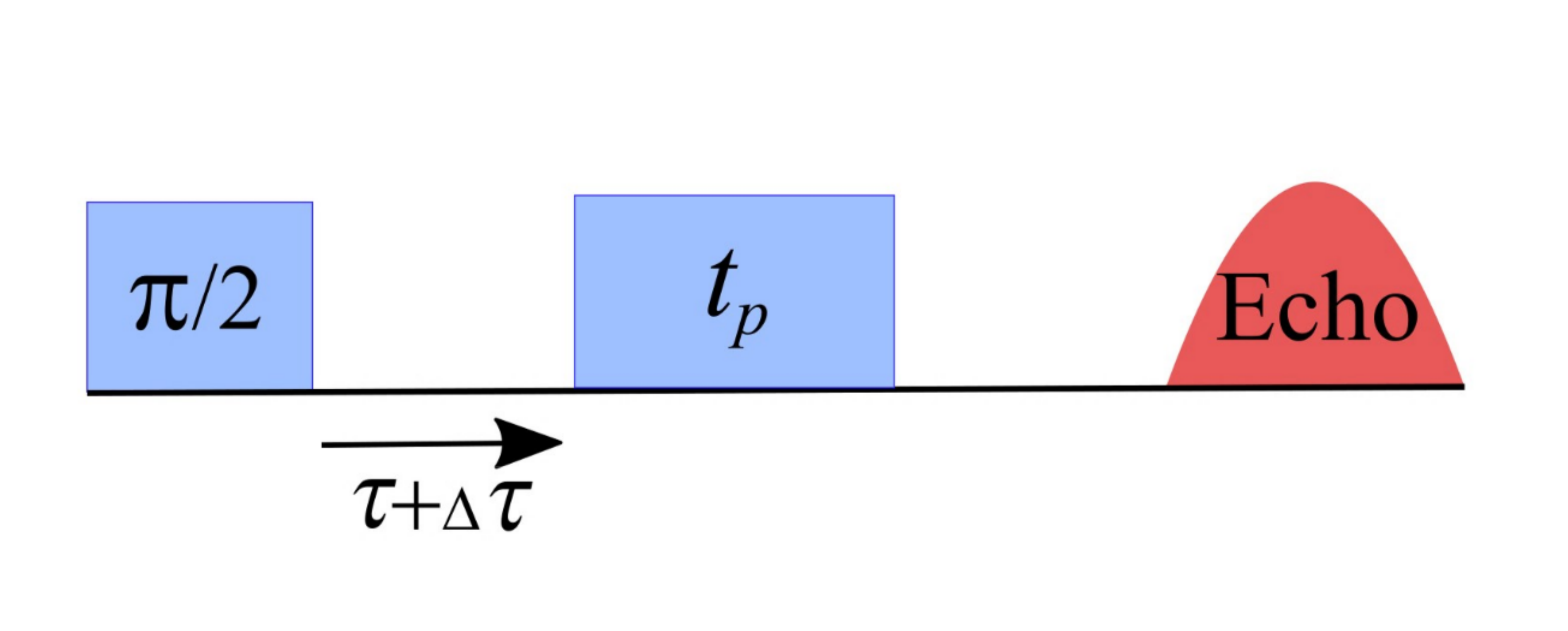}
	\caption[]{Hahn echo sequence for pulsed EPR measurements. The sequence allows measuring  (by varying $\tau$) the $T_2$ time. To probe instantaneous diffusion,  $T_2$ time measurements were repeated for different lengths of the refocusing pulse ($t_p$) at constant microwave power. }
	\label{fig:T2}
\end{figure}

For the $^{\mathrm{NV}}T_2$ measurements, we used the standard Hahn echo sequence, that is a $\pi/2$ followed by a refocusing ($\pi$) pulse, as shown in Fig.~\ref{fig:T2}.

\begin{figure}[htb!]
	\centering
	\includegraphics[width=3.5in]{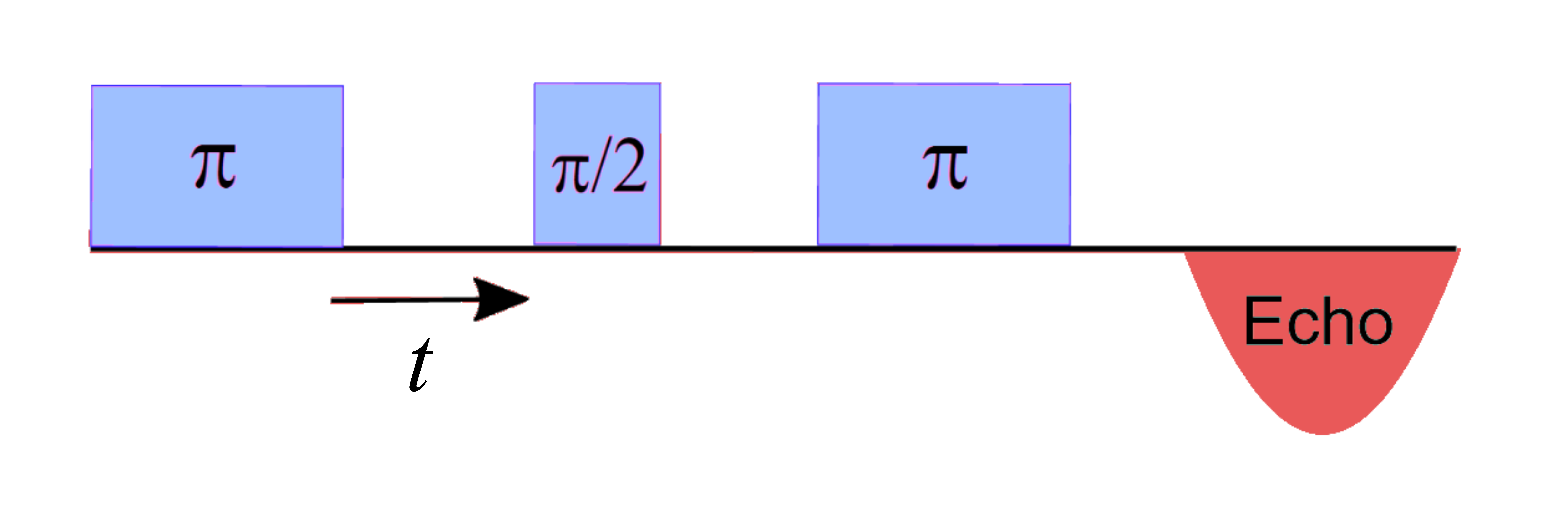}
	\caption[]{Inversion-recovery sequence for $T_1$ measurements.}
	\label{fig:T1}
\end{figure}

To measure $^{\mathrm{NV}}T_1$ within the \SI{2}{\micro\meter} particles (main text, Table~\ref{table2}), we used the Inversion-Recovery sequence (Fig.~\ref{fig:T1}), that consists in applying first an $\pi$ (inversion) pulse, waiting a time $t$, and then performing readout (readout was done with the Hahn echo sequence). $T_1$ values were extracted by fitting the recovery curve, or the signal as a function of $t$. The recovery curve was fitted with a biexponential function. The short component ($\sim$~\SI{100}{\micro\second}) was attributed to spectral diffusion, that occurs because of the limited bandwidth of the inversion pulse, while the long component was attributed to $^{\mathrm{NV}}T_1$.

Inversion-recovery was found not to be a viable method for the 100 nm particles. There, the intrinsic $T_1$ relaxation and spectral diffusion occur with comparable rate, which means both mechanisms cannot be conveniently distinguished with a biexponential fit. In order to eliminate spectral diffusion, we resorted to a \textit{saturation-recovery} sequence, by replacing the inversion pulse with a train of $\pi/2$ pulses (saturation). Using $n=25$  pulses spaced by \SI{500}{\micro\second}, we observed a single exponential recovery. Fitting such a recovery curve gives yields the $^{\mathrm{NV}}T_1$ values given in main text, Table~\ref{table100nm}.

\section{Pulsed EPR Spectrum}
The spectra in Figure \ref{fig:spec} of the main text were acquired by measuring the echo amplitude of a Hahn echo sequence ($\pi/2$-pulse + waiting time $\tau$ + $\pi$-pulse) as a function of the static magnetic field $H$. The accumulated intensities in Figure \ref{fig:spec} are affected by irreversible decoherence processes ($T_2$) that occur during the waiting time between pulses in Hahn echo sequence. Using the $T_2$ times for NV$^-$ and P1 we can estimate the signal loss factor ($\alpha$) for each of the spectral intensities:  \\
\begin{equation}
\alpha=\exp(0)-\exp\left(-\frac{2\tau+t_{\pi}}{T_2}\right),
\end{equation}
where $\tau$ is the waiting time between the $\pi/2$ and refocusing ($\pi$)   pulse in the Hahn echo sequence and $t_{\pi}$ is the length of the $\pi$ pulse (see Figure~\ref{fig:T2}). We used $\tau=120$~ns and $t_{\pi}=60$~ns.   \\
 \indent The signal experiences a loss depending on $T_2$ and $\tau$ meaning that the P1 spectrum, which has a significantly shorter $T_2$ than NV$^-$, experiences a greater loss in spectral intensity. However, this mechanism does not result in any important difference in the attenuation factor  for the same center (see Table \ref{tableS1}). Therefore, the main reason for the observed significant change in the NV$^-$/P1 line intensities is the increased conversion efficiency that results in higher concentration of NV$^-$ and lower concentration of P1.
\textbf{\begin{table}[htb!]
		\caption{The respective signal intensity losses for the ESE-EPR spectra shown in Figure \ref{fig:spec}.}
		\begin{adjustwidth}{-.5in}{-.5in} 	
			\begin{center}
				\begin{tabular}{ l l l l l}
					\hline
					\thead {Defect \\ ~} &  \thead  {Sample name\\ ~}   & \thead  {Hahn echo \\  $T_2$ time ($\mathbf{\SI{}{\micro\second}}$ )}   & \thead  {Signal loss \\ (\%)} & \thead  {$T_2$-induced  \\ signal variation (\%)} \\ 
					\hline
					 \multirow{2}{*}{NV$^-$} & \multicolumn{1}{l}{0.5MSY2} & \multicolumn{1}{l}{2.2} & \multicolumn{1}{l}{12.7}   & \multirow{2}{*}{-0.7}  \\ 
					 &\multicolumn{1}{l}{6MSY2} & \multicolumn{1}{l}{2.1} & \multicolumn{1}{l}{13.3} \\ 
				     \multirow{2}{*}{P1} & \multicolumn{1}{l}{0.5MSY2} & \multicolumn{1}{l}{1.2}& \multicolumn{1}{l}{22.1}  & \multirow{2}{*}{+4.1}  \\ 
				    &\multicolumn{1}{l}{6MSY2} & \multicolumn{1}{l}{1.4}& \multicolumn{1}{l}{18.9} & \\ 
				    \hline	
				\end{tabular}
			\end{center}
			\label{tableS1}
		\end{adjustwidth} 
\end{table}}

\clearpage
\section{Photoluminescence spectra}
The photoluminescence (PL) spectra for the 0.5MSY2, 3MSY2 and 6MSY2 samples were collected at room temperature, each from a single diamond particle. The measurement have been implemented using a home-built confocal setup (see  Experimental section in the main text), equipped with the spectrometer (Princeton Instruments, IsoPlane 160 with a camera Pixis 100B) and applying the laser power of $\SI{3.5}{\micro\watt}$.
\begin{figure}[htb!]
	\centering
	\includegraphics[angle=270,width=4in]{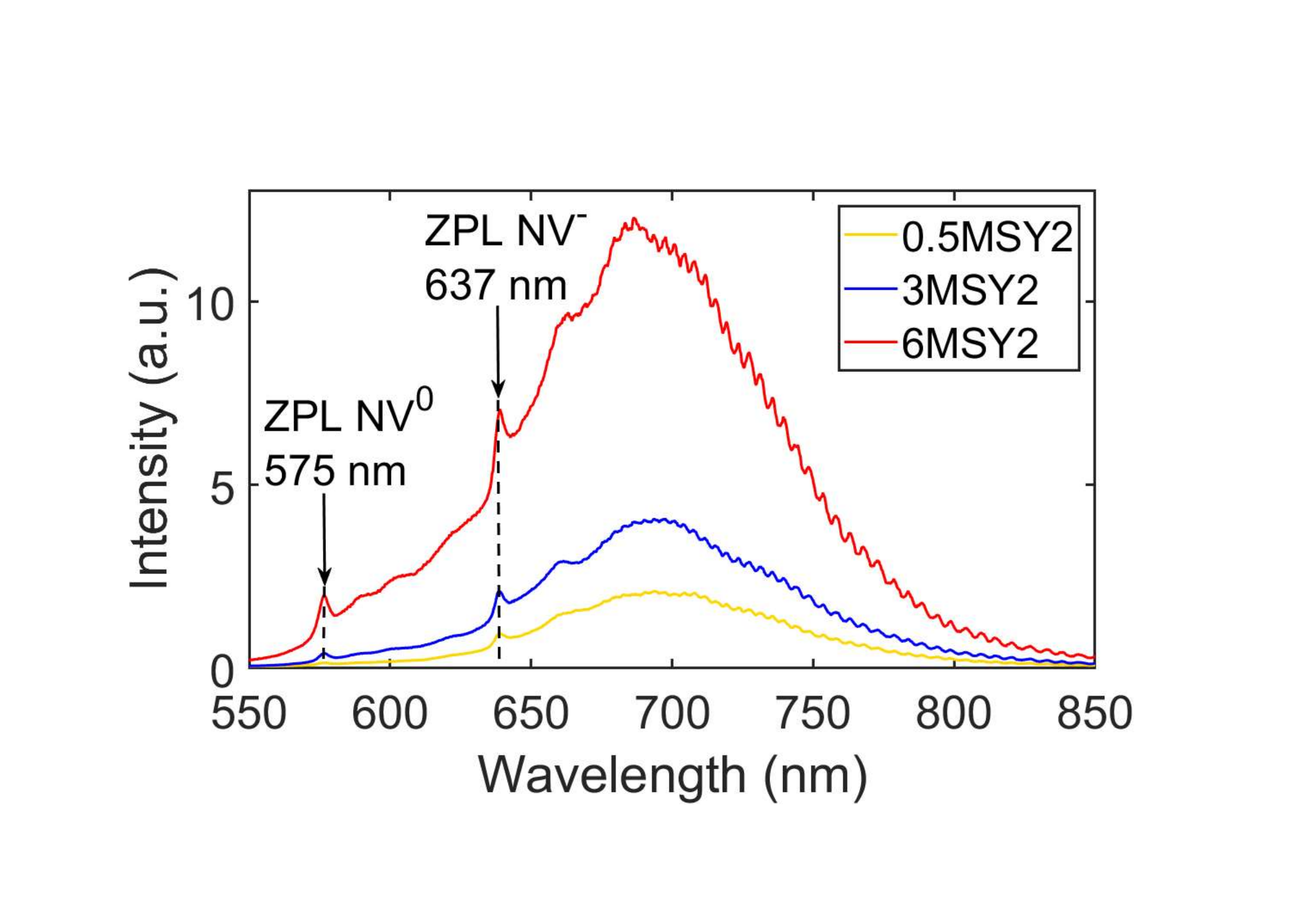}
	\caption[]{PL spectra of the 0.5MSY2, 3MSY2 and 6MSY2 samples. We can observe the NV$^0$ and NV$^-$ centers' zero-phonon lines (ZPL) and their broad phonon sideband. The excitation wavelength is \SI{532}{\nano\meter}.}
	\label{fig:PL}
\end{figure}

The increase in the PL intensity   corresponding to NV$^-$, seen in Figure \ref{fig:PL}, might originate from the raise in NV$^-$ concentration with increasing  irradiation dose, as we discuss in the section "Effects of increasing the irradiation dose" in the main text. However, the PL intensity is affected by other factors. In particular, for $\SI{2}{\micro\meter}$ powder  scattering effects and  interference within single particles may lead to different PL signal and spectral distortions, depending on the particle shape. For the precise interpretation of PL spectra for the $\SI{2}{\micro\meter}$ diamond particles, further work, particularly the modeling of scattering effects, would be needed.

\clearpage
\section{Instantaneous Diffusion}

Alternatively to the CW method, the evaluation of the spin density can be implemented with Pulse EPR based on the effect of instantaneous diffusion (ID). In contrast to CW EPR, the ID method probes the dipolar interaction between the target spins, which decreases rapidly with distance. It provides information about the local spin density in the sample. As a consequence, it does not require a spin standard or any knowledge of the sample mass.

We here present the results of ID measurements on the  \SI{2}{\micro\meter} samples irradiated at high temperature with different doses (described in the main text, section ``Effects of increasing the irradiation dose''). Determination of the P1 concentration was achieved for all these HT  irradiated \SI{2}{\micro\meter} samples (concentrations in  Table~\ref{table2}), and is discussed  below in section ``Analysis of Instantaneous Diffusion for P1 centers''. 
Evaluating the NV$^-$ concentration through ID is possible only for samples in which decoherence from other sources (e.g. spectral diffusion induced by P1) do not play an exceedingly important role. In our case, this was achieved for the samples with the highest conversion efficiencies, as the number  of closely interacting NV$^-$-NV$^-$ pairs is then  sufficiently important, and the effect of the P1 spin bath is reduced (see also main text, ``Non-NV$^-$ irradiation-induced defects''). The analysis method, and the estimate of NV$^-$ from ID for one sample, 6MSY2, are discussed below in section ``Analysis of Instantaneous Diffusion for NV$^-$ Centers''.
The case of the \SI{100}{\nano\meter}  particles would require further investigations.

Practically, the ID  experiment consists of measuring the Hahn echo decay for different flip angles $\theta$ of the refocusing pulse, which we perform by varying the length of this pulse at constant microwave power (Figure~\ref{fig:T2}). Usually, the concentrations are extracted from the linear fit of the experimental data \cite{Lund17,ID2}: 
\begin{equation}
1/T_2=A+B\left<\sin^{2}{(\theta/2)}\right>\mathrm{,}
\label{eq.1}
\end{equation}
\noindent where $\left< \right>$ denotes an average over all possible flip angles, $A$ corresponds to $1/T_2$ in the absence of the ID effect, \cite{Tyryshkin} and:
\begin{equation}
B=4\pi^{2}\gamma^{2}\hbar C/9\sqrt{3}= C\cdot 8.2\times 10^{-13} ~\mathrm{cm}^3 \mathrm{s}^{-1}\mathrm{.}
\label{eq.2}
\end{equation}

However, this description was originally made  for a system of  1/2 spins \cite{salikhov1981}. As a consequence, Eq.~\eqref{eq.1} and Eq.~\eqref{eq.2} need to be adapted for NV$^{-}$, which is a spin-1 system.

\subsection*{Analysis of Instantaneous Diffusion for NV$^-$ Centers}

\begin{figure}[htb!]
	\centering
	\includegraphics[angle=270,width=4in]{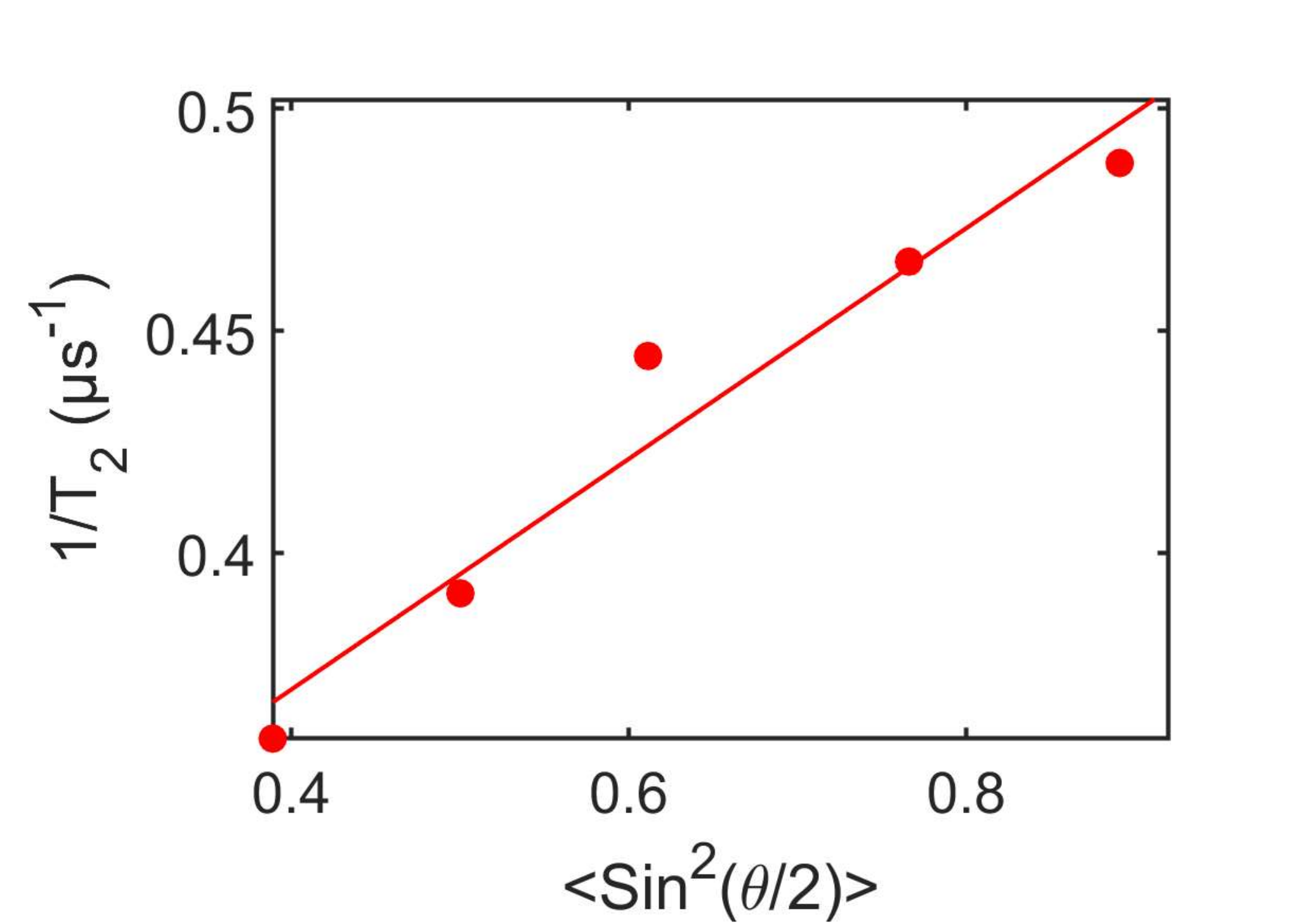}
	\caption[]{The dependence of $\mathit{^{\mathrm{NV}}T_2}$ of NV$^-$ centers on the flip angle for the 6MSY2 sample. The solid line is linear regression.}
	\label{fig:ID_NV}
\end{figure}

 Instantaneous diffusion allowed to determine the NV$^-$ concentration in the 6MSY2 sample (see Figure \ref{fig:ID_NV}). In this case, $T_2$ decays were measured for different lengths of the refocusing pulse on the low-field most intensive line of the Pake doublet (at 2870 G magnetic field).

When analyzing the experimental data, it is important to consider again that instantaneous diffusion depends on the interaction between paramagnetic spins on a local scale, i.e. between a few of the nearest neighbours of the probed species. 

Table \ref{tab:NNdist} shows the average  nearest neighbour distances between paramagnetic impurities assuming concentrations in the range corresponding to NV$^-$   in our powder samples (1-100 ppm).  
For such a dilute system, the nearest-neighbour distribution function obeys to a Poissonian statistic with an average distance \mbox{$\left\langle r_{\mathrm{NN}} \right\rangle = 0.55 C^{-1/3}$}, where $C$ is the concentration per unit volume \cite{reynhardt2001}. The obtained values for $\left\langle r_{\mathrm{NN}} \right\rangle$ are below 10 nanometers (Table~\ref{tab:NNdist}). 
We now consider the case in which the lengthscale probed by instantaneous diffusion is inferior to the particle size, which is the case for the  powder samples for which ID results are shown (\SI{2}{\micro\meter}). On the small lengthscale over which the interaction between spins occurs, we expect the material to be monocrystalline, that is the NV$^-$ axis can have only four possible orientations with respect to the external magnetic field.

\begin{table}[htb!]
	\caption{Average nearest-neighbour distances calculated for a few concentrations 
		typical for the NV$^-$ defects in our samples (see text).}
	\begin{center}
		\begin{tabular}{c c c}
			\hline
			\thead {Concentration \\ (ppm)}   &  \thead  {Concentration \\ (cm$^{-3}$)} & \thead {Average nearest-neighbour \\   distance $\left\langle r_{\mathrm{NN}} \right\rangle $     (nm) } \\  
			\hline
			1 &  $1.76\times 10^{17}$  & 9.8      \\    
			
			10  & $1.76\times 10^{18}$  & 4.5     \\    
			
			100  & $1.76\times 10^{19}$  & 2.1     \\    
			\hline
		\end{tabular}
	\end{center}
	\label{tab:NNdist}
\end{table}

To interpret the instantaneous diffusion data shown in Figure~\ref{fig:ID_NV} for such particles, 
one needs therefore to consider that such a local (monocrystalline) order exists within each diamond particle.

\begin{figure}[h!]
	\centering
	\includegraphics[width=5in]{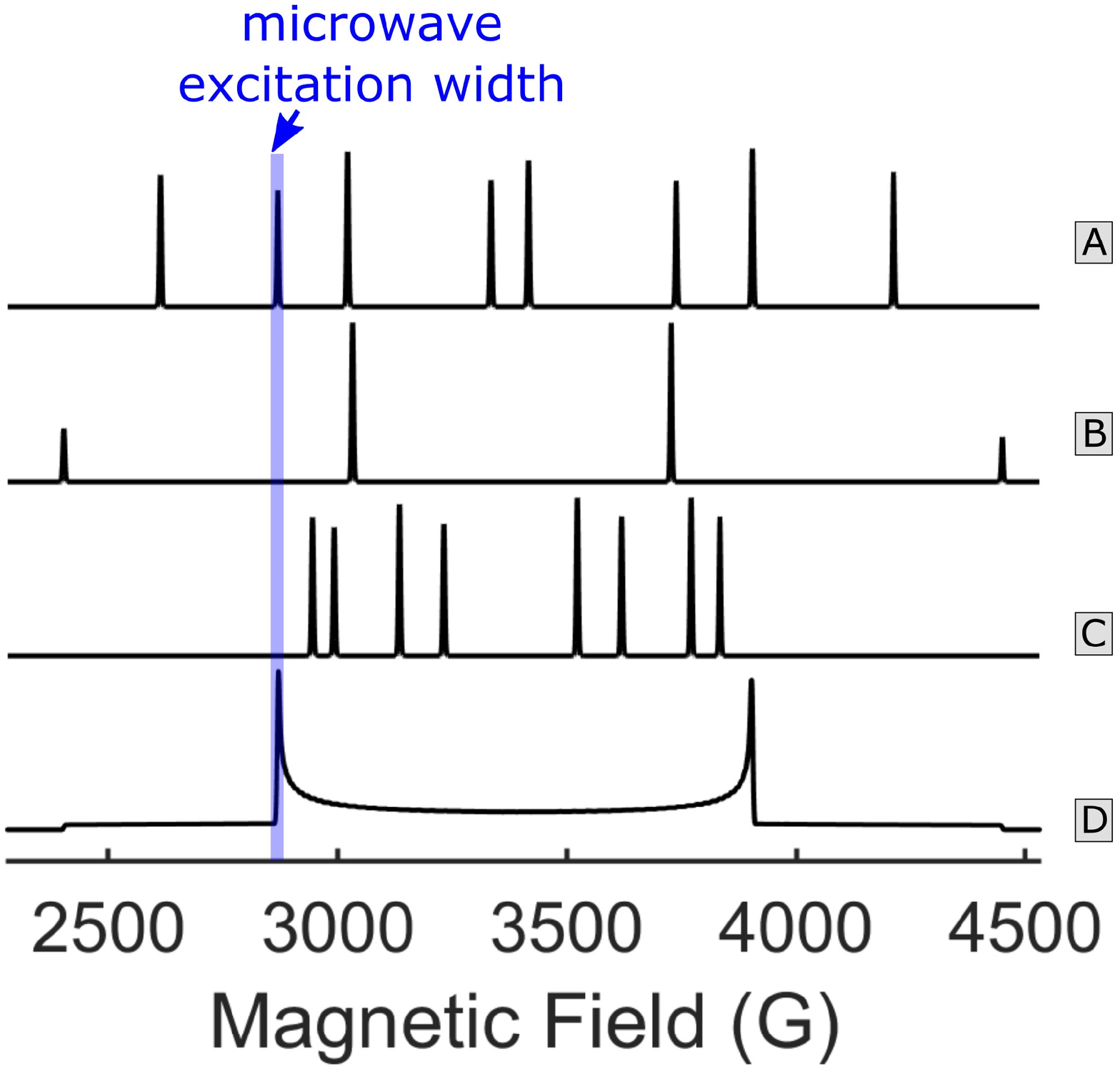}
	\caption[]{Simulation of EPR absorption NV$^-$ spectra from single crystal (A,B,C) and powder (D), at frequency $f=9.6$~GHz.
		Single crystal spectra shown here correspond to a few selected orientations of  the crystal with respect to the magnetic field $\mathbf{B_0}$: A: $\mathbf{B_0}$  along the direction $\mathbf{u_A}=[-0.35 \ -0.65 \ 1]$, that is perpendicular to the $[1\ 1\ 1]$ crystallographic direction; B: $\mathbf{B_0}$ \emph{parallel} to $[1\ 1\ 1]$; C: $\mathbf{B_0}$ in an arbitrary direction, along  $\mathbf{u_C} = [0.34\ 0.12 \ 1]$. 
		The powder spectrum (D) corresponds to the accumulation of single crystal spectra for all possible orientations.  The vertical shaded area highlights the excitation width of the refocussing microwave pulse if the field is set to the 2870 G peak. Only particles with certain orientations (e.g., A) exhibit NV$^-$ resonant with the microwave, while no resonance occur for powder particles/single crystals with other orientations (such as B and C).
	}
	\label{fig:NVsim}
\end{figure}

A few simulated single crystal EPR spectra are shown in Figure~\ref{fig:NVsim}. Each spectrum shows 8 spectral lines, as each of the 4 possible orientations of NV$^-$ gives two lines (resulting from the two spin-1 transitions).
The position of these lines depends strongly on the crystal orientation  with respect to the magnetic field, which means that the resonance between NV$^-$ and the microwave excitation occurs only for specific powder grain orientations, as we discuss now.

\begin{figure}[h!]
	\centering
	\includegraphics[width=290pt]{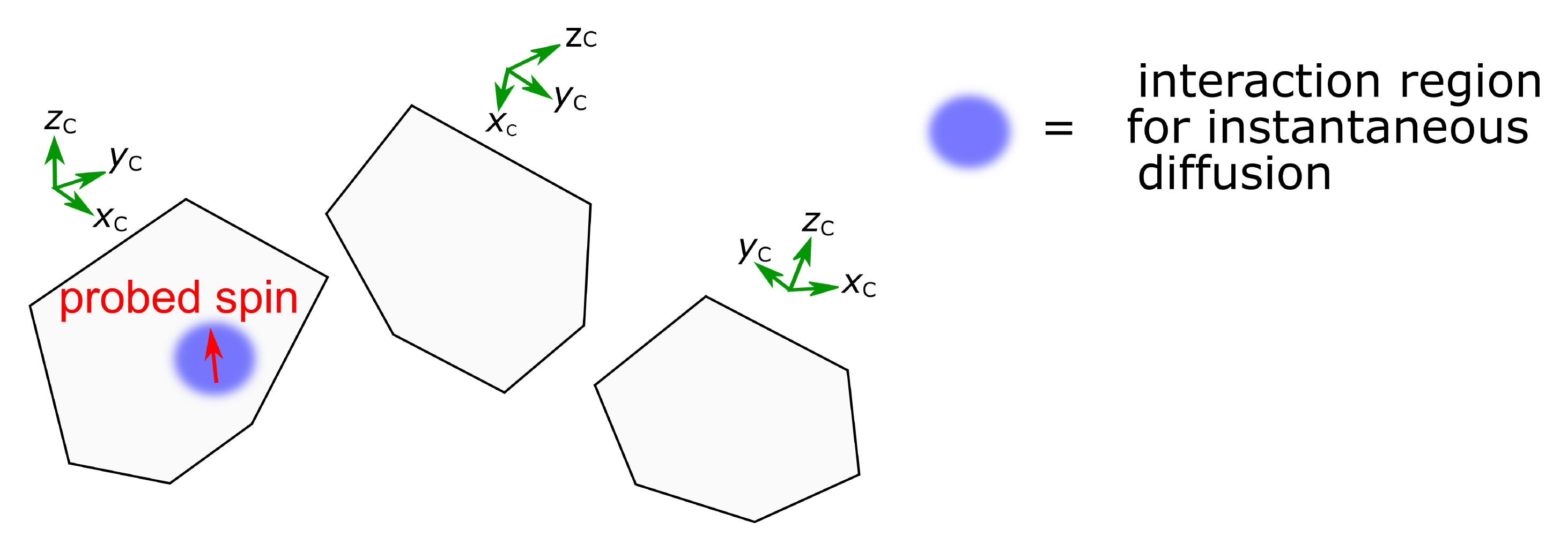}
	\caption[]{Description of the instantaneous diffusion experiment in micrometric powder samples. The experiment relies on the interaction between the first nearest neighbour spins, which occur in a volume (blue region) smaller than the grain volume. For a given microwave frequency, some powder grains will not contain resonant spins (see text).}
	\label{fig:ID_powder_grains}
\end{figure}

The microwave excitation is effective within a range of frequencies $\pm f_a$, where $f_a$ is the Rabi frequency of the microwave field, here 8.3~MHz. We convert this quantity to the field scale by dividing by the gyromagnetic ratio \mbox{$\gamma = $ \SI{2.80}{\mega\hertz\per\gauss}}, which gives $H_a = f_a / \gamma \sim 3$~G. The excitation width of $2 H_a = 6$~G is very small in comparison to the width of the field range where NV$^-$ transitions can appear, that is $2 D / \gamma = 2050$  G (where $D$ is the zero-field splitting of NV$^-$). As illustrated in Figure~\ref{fig:ID_powder_grains}, although powder grains with specific orientations will respond to the microwave excitation (spectrum A in  Figure~\ref{fig:NVsim}), some other orientations will remain unaffected (spectra B and C, in Figure~\ref{fig:NVsim}).  Furthermore, even for powder particles of specific orientation that contain NV$^-$ resonant with the microwave, we expect the resonance to typically occur for a single spectral line (the overlap of two or more lines within the narrow  excitation bandwidth is very rare).

Therefore, the existence of a local orientational order means, that an ID experiment  for  NV$^-$ in powder particles  that are  sufficiently big in comparison to the average nearest neighbour distance (Table   \ref{tab:NNdist}), $\gtrsim$   \SI{100}{\nano\meter},   should be analyzed the same way as   single crystal experiments (for which the crystal is oriented so that only one spectral line is resonant with the microwave excitation).
We remark again that Eq.~\eqref{eq.1} and Eq.~\eqref{eq.2} correspond to a description that was made for a spin 1/2 system  \cite{salikhov1981}, and therefore do not directly apply to NV$^-$. 
We now discuss the adaptations that need to be made to these formula to recover the full NV$^-$ concentration. 

\begin{figure}[h!]
	\centering
	\includegraphics[width=2.5in]{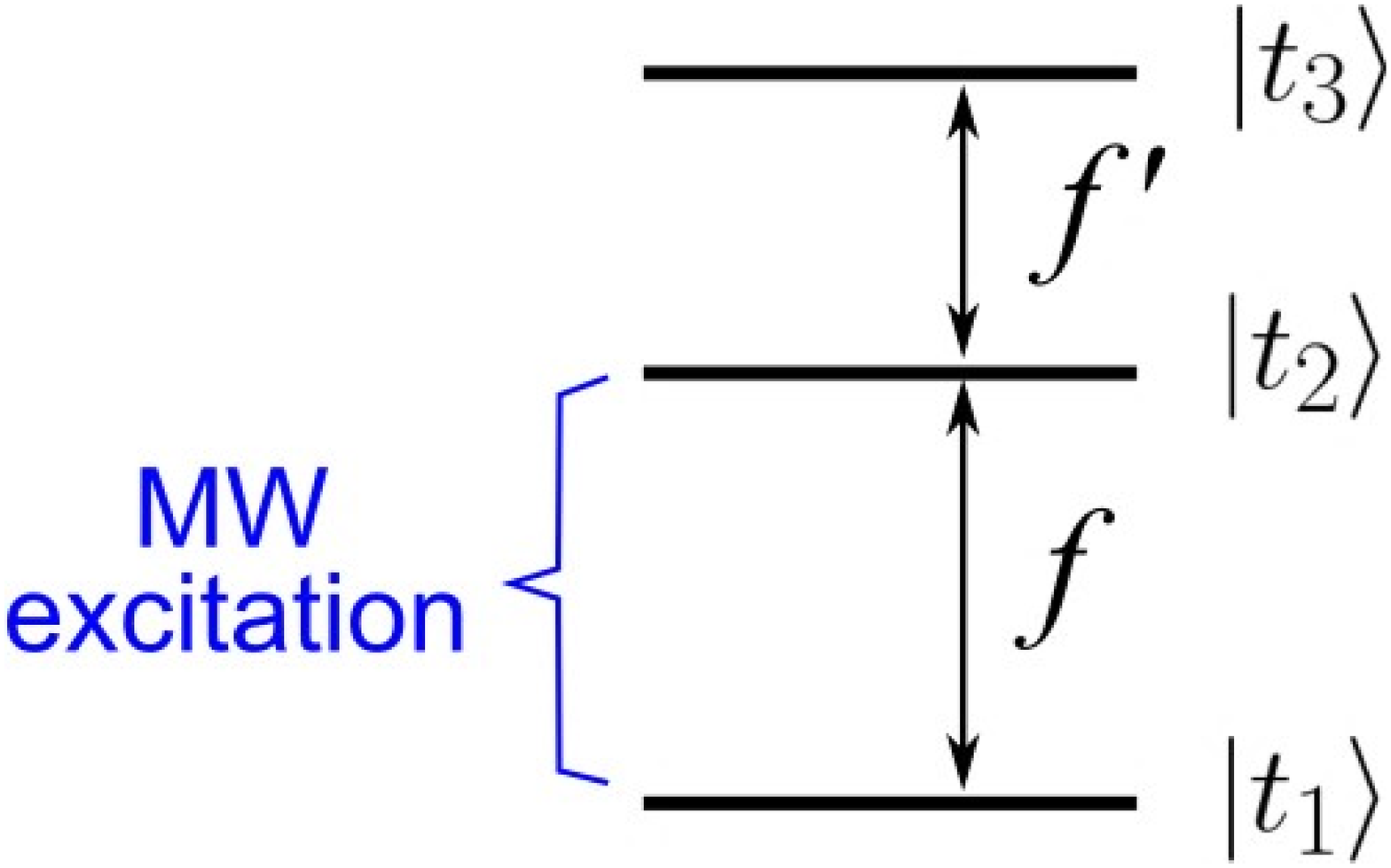}
	\caption[]{Energy diagram for NV$^{-}$, 
		with the microwave (MW) excitation resonant to only one transition. }
	\label{fig:NV_en_diagram}
\end{figure}

For this, we consider a single crystal and label the four NV$^{-}$ orientations $\mathbf{v_a}$, $\mathbf{v_b}$,  $\mathbf{v_c}$, $\mathbf{v_d}$, and assume only the NV$^{-}$ with orientation  $\mathbf{v_a}$ is resonant with the microwave. For this particular orientation, we label $\ket{t_i}, i=1,2,3$ the energy eigenstates of NV$^{-}$. The experiment described in the present work was performed at a magnetic field of 2870 G and frequency $f=9.6$~GHz on resonance with the low-field intensive line of the Pake doublet ($\ket{t_1} \leftrightarrow \ket{t_2}$). In that case, the simulation predict that the other transition appears at $f' = 6.74$~GHz, which is well separated in frequency, and will therefore not be resonant with the microwave (Figure~\ref{fig:NV_en_diagram}).

To participate in instantaneous diffusion, one NV$^{-}$ needs not only to be  in one particular orientation ($\mathbf{v_a}$), but also to be in one of the two spin states involved in the transition ($\ket{t_1}$ or $\ket{t_2}$), at the time of the measurement. Therefore, analyzing the experimental data with Eq.~\eqref{eq.1} yields only the concentration for a \textit{subset} of NV$^{-}$, that are in the correct orientation and spin state. 
We remark that, for this subset of NV$^{-}$, the flip angle $\theta$ in Eq.~\eqref{eq.1} is proportional to the length of the refocusing pulse, $\theta=\pi t_p/t_{\pi}$ (where $t_{\pi}$ is the duration of 
the full $\pi$ pulse).  To  recover the \textit{total} concentration of NV$^{-}$  $C_{\mathrm{tot}}$, one must  renormalize the $C$ value obtained by fitting the experimental data with Eq.~\eqref{eq.1} and Eq.~\eqref{eq.2}, with the following formula: 

\begin{equation}
\label{eq.ID_prefactor}
C_{\mathrm{tot}}  =  \frac{4}{\rho_1 + \rho_2} C,
\end{equation}

\noindent where $\rho_1, \rho_2$ correspond to the occupation probability of states $\ket{t_1}$ and $\ket{t_2}$, respectively. Multiplication by 4 accounts for the selectivity of the measurement in terms of NV$^{-}$ orientation, while division by $(\rho_1 + \rho_2)$ accounts for its selectivity in terms of initial spin state.
For a precise estimate of the prefactor in Eq.~\eqref{eq.ID_prefactor}, one can calculate $\rho_1$ and $\rho_2$ from the thermal populations at a given temperature $T$:
\begin{align*}
\rho_1   &=  \frac{1}{1 + e^{-hf/k_B T} +e^{-h(f+f')/k_B T}}, \ \textrm{and :} \\
\rho_2   &=  \frac{e^{-hf/k_B T} }{1 + e^{-hf/k_B T} +e^{-h(f+f')/k_B T}},
\end{align*}

\noindent where $k_B$ and $h$ are the Boltzmann and Planck constant respectively. 
At room temperature, $T=293$~K, we obtain: 

\begin{equation}
\label{eq.ID_prefactor.2}
C_{\mathrm{tot}}  = 5.996 \ C \approx 6 \ C.
\end{equation}

The measurements were carried out by exciting the intensive line corresponding to NV$^-$ spins oriented perpendicular to the magnetic field, at $\sim \SI{2900}{\gauss}$ and the received result was multiplied by the the factor in Eq.~\eqref{eq.ID_prefactor.2}. The received NV$^-$ concentration for the 6MSY2 sample is 10.8 ppm, which is compatible with the result obtained separately with CW EPR (10.3~ppm).

Finally, by considering the offset in the linear fit shown  in Figure~\ref{fig:ID_NV}, one can obtain the value of $T_2$ \emph{in the absence of instantaneous diffusion}, which is \SI{3.8}{\micro\second}. The important increase in  $T_2$   in comparison with the outcome of the   standard Hahn-echo measurement  (\SI{2.1}{\micro\second}), confirms that NV$^-$-NV$^-$ interaction plays a sizeable role in decoherence.  In addition, the  interaction between NV$^-$ can also be a factor for decoherence through  spectral diffusion, an effect which was not measured here.

\subsection*{Analysis of Instantaneous Diffusion for P1 Centers}

\begin{figure}[h!]
	\centering
	\includegraphics[width=5in]{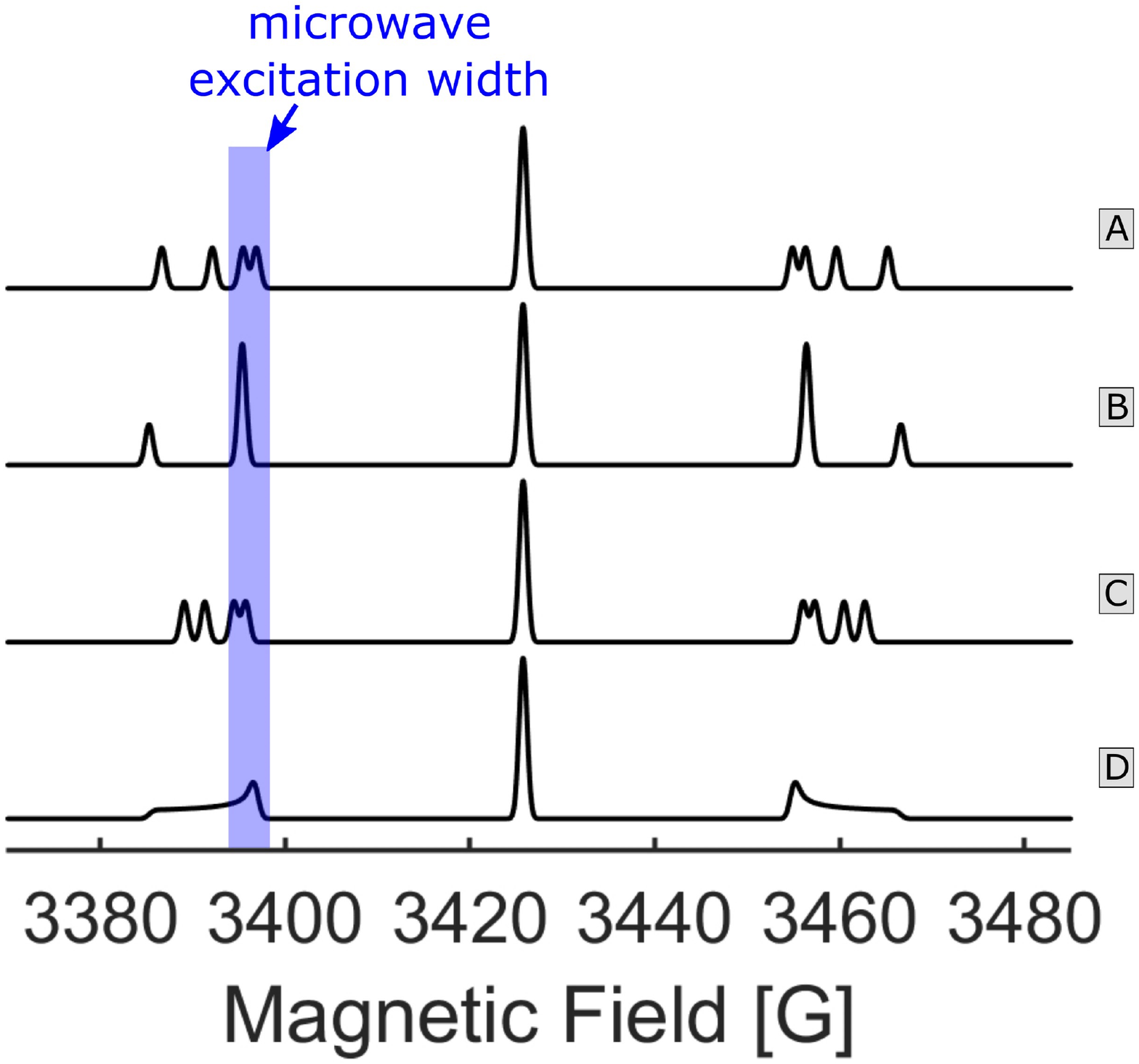}
	\caption[]{Simulation of EPR absorption P1 spectra from single crystal (A,B,C) and powder (D), at frequency $f=9.6$~GHz.
		As in Figure~\ref{fig:NVsim}, single crystal spectra shown here correspond to a few selected orientations of  the crystal with respect to the magnetic field $\mathbf{B_0}$: A: $\mathbf{B_0}$  along the direction $\mathbf{u_A}=[-0.35 \ -0.65 \ 1]$, that is perpendicular to the $[1\ 1\ 1]$ crystallographic direction; B: $\mathbf{B_0}$ \emph{parallel} to $[1\ 1\ 1]$; C: $\mathbf{B_0}$ in an arbitrary direction, along  $\mathbf{u_C} = [0.34\ 0.12 \ 1]$. 
		The powder spectrum (D) corresponds to the accumulation of single crystal spectra for all possible orientations.  The vertical shaded area highlights the excitation width of the refocussing microwave pulse if the field is set to the maximum of the P1 low-field hyperfine line, demonstrating that more than one resonance occur within the excitation bandwidth of the refocusing pulse.}
	\label{fig:P1sim}
\end{figure}

The P1 concentrations  were measured with the instantaneous diffusion effect on the maximum of the low-field hyperfine line. The P1 spectrum is less anisotropic than the NV$^-$ spectrum. Therefore, the bandwidth of the refocusing pulse in the sequence for measuring $T_2$ (Figure \ref{fig:T2}) covers a bigger area, affecting P1 centers with different orientations (see Figure \ref{fig:P1sim}). In this case the shape of the P1 spectrum has to be considered when calculating the term that depends on the flip angle ($\left<\sin^{2}{(\theta/2)}\right>$)  in Equation~\eqref{eq.1}. This is done by applying the following formula \cite{T2_field}:

\begin{equation}
\left<\sin^{2}{(\theta/2)}\right>=\frac{\int\text{d}H f(H)\frac{H_{a}^2}{(H-H_{res})^2+H_{a}^2}\sin(\frac{\gamma t_{p}}{2}\sqrt{(H-H_{res})^2+H_{a}^2})}{\int\text{d}Hf(H)},
\label{inst}
\end{equation}

\noindent where $f(H)$ describes the P1 spectrum; $H_{a}$ is the amplitude of the microwave field; $H_{res}$ the resonance magnetic field; $t_{p}$ the duration of the refocussing pulse and $\gamma$ the gyromagnetic ratio.

\begin{figure}[htb!]
	\centering
	\includegraphics[width=4in]{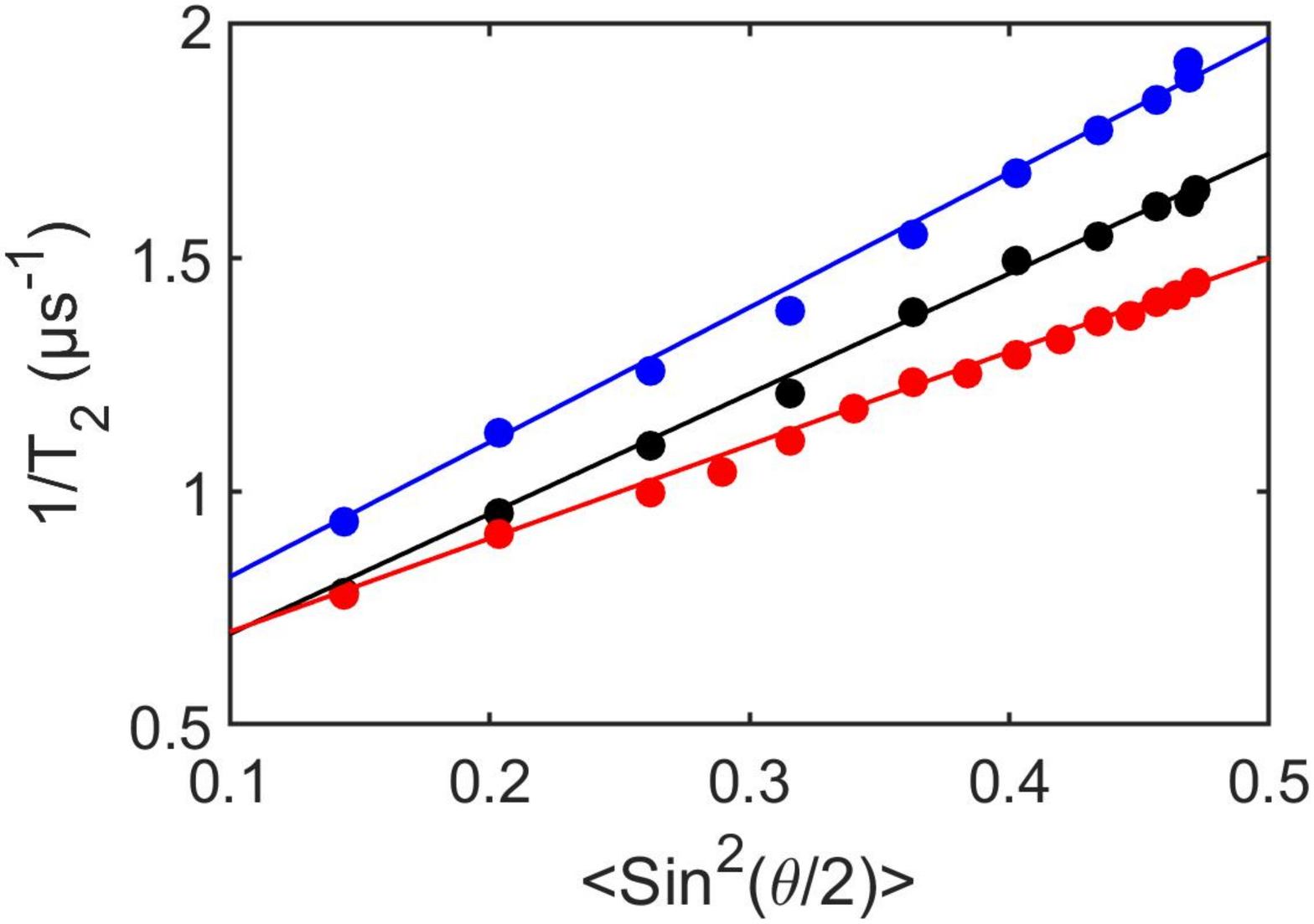}
	\caption[]{(a) Dependences of $\mathit{^{\mathrm{P1}}T_2}$ of P1 centers on the flip angle ($\theta$) caused by instantaneous diffusion for the 0.5MSY2 (blue),  3MSY2 (black) and 6MSY2 (red) samples. The solid lines are linear regressions.}
	\label{fig:ID_P1}
\end{figure}

The experimental data of $T_2$ using different lengths of $\pi$ pulses for P1 centers fitted with the Equation \eqref{eq.1} for the 0.5MSY2, 3MSY2, 6MSY2 samples is  presented in Figure~\eqref{fig:ID_P1}.  The obtained concentrations for all the HT irradiated samples  $\mathbf{\SI{2}{\micro\meter}}$ samples (described in main text section ``Effects of increasing the irradiation dose'') are given as well here in Table~\ref{table_CWvsID_P1}.

\begin{figure}[h!]
	\centering
	\includegraphics[width=4in]{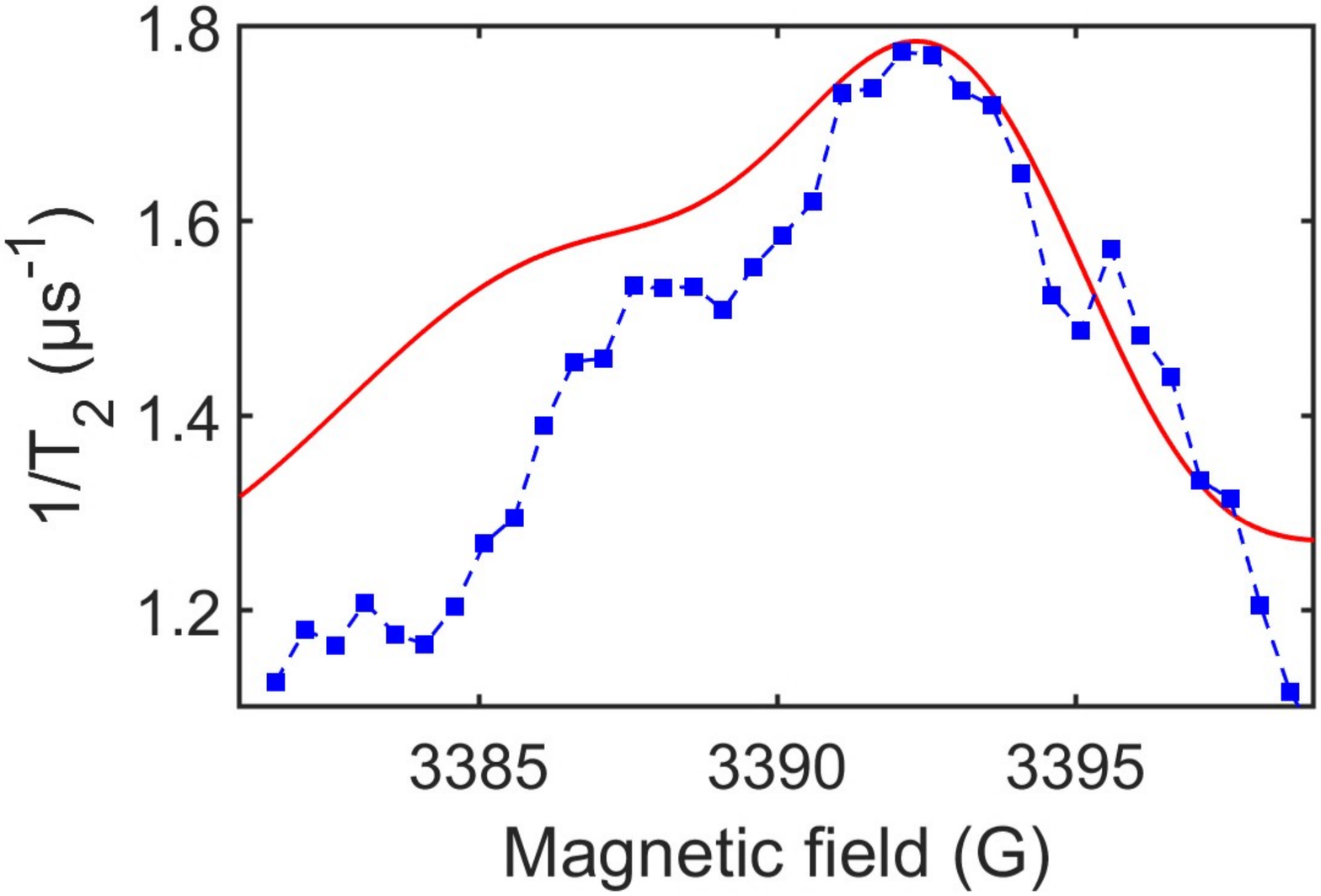}
	\caption[]{Dependence of the $\mathit{^{\mathrm{P1}}T_2}$ on the magnetic field position for an unirradiated \SI{2}{\micro\meter} sample, caused by instantaneous diffusion: the blue squares represent the experimental data. The blue dashed line is a guide for the eye. The red solid line is the simulation. }
	\label{fig:T2_field}
\end{figure}

Additionally, since the value of $T_2$ is dependent on the density of spins affected by the refocusing pulse, the shape of the spectrum can be reproduced by measuring the decay rate $1/T_2$ for different field positions~\cite{T2_field}. The effect of ID on the decay rate ($1/T_2$) gives an indication of the amount of spins in the spin packet affected by the $\pi$ pulse. Keeping the length of the pulses for the $T_2$ measurement fixed, but changing the magnetic field, different spin packets will contribute to the $T_2$ decay, repeating the shape of the spectral line. Knowing the concentration of defects and using Equation \ref{inst} it is possible to reproduce the decay rate, as a function of field position. The experimentally received values of $1/T_2$ for different magnetic field positions for the P1 hyperfine line for the unirradiated \SI{2}{\micro\meter} sample (0MSY2)  and the respective simulation using Equation \ref{inst} are presented in Figure \ref{fig:T2_field}.

From Table \ref{table_CWvsID_P1}, we  observe overall a good agreement between the concentrations obtained with ID and CW spin-counting for the HT irradiated \SI{2}{\micro\meter} samples up to dose \SI{3e18}{\per\centi\meter\square}. For the samples with the highest irradiation doses (6MSY2 and 9MSY2), ID provides a significantly higher result in comparison with CW EPR. The reason behind such discrepancy is not established at the moment, but could potentially reveal an inhomogeneous P1 distribution in the samples irradiated at high doses.

\begin{table}[htb!]
	\caption{Comparison of P1 concentrations obtained with CW and instantaneous diffusion for the HT irradiated \SI{2}{\micro\meter} samples.}	
	\begin{center}
		\begin{tabular}{  c  c c }
			\hline 
			\thead{Sample\\name}&   \thead  {P1 conc., \\ CW (ppm) } & \thead {P1 conc., \\ ID (ppm)}   \\ 
			\hline
	    	0MSY2  & 53  & 64   \\ 
			0.5MSY2  & 55   & 60  \\ 
			1MSY2  &   41  & 53   \\
			2MSY2  &   48  & 53   \\ 
			3MSY2  &   40  & 53  \\
			6MSY2 &   18  & 41   \\   
			9MSY2 &   13  & 30   \\   \hline	
		\end{tabular}
	\end{center}
	\label{table_CWvsID_P1}
\end{table}

Finally, for smaller samples (\SI{100}{\nano\meter}), the analysis of ID might need to be adapted, as we detail now. 
From the CW spectra (not shown), it can be seen that the hyperfine line of P1  overlaps with a broad $g \approx 2$ spins contribution that was attributed to deformation-related defects at the vicinity of the surface (e.g., labelled SC1 in Ref.~\cite{yavkin2015}). Unlike in the \SI{2}{\micro\meter} case, such overlap is significant in the case of the \SI{100}{\nano\meter} samples, as a  consequence of the increase in surface-to-volume ratio. Because of this overlap, the ID effect could potentially be modified as a consequence of the interaction of P1 with the spins that provide this broad spectral  component. Thus, the analysis of ID data for NDs of \SI{100}{\nano\meter} size  would probably require examining whether the  interaction of these additional spins with P1 centers  plays a role in  the ID effect, and, if yes, how it would affect the concentration determination for P1.

\clearpage
\bibliography{YM_HT_irradiation_Supp_Inf_Carbon}